\documentclass[a4paper,11pt]{article}

\usepackage{jcappub} 

\usepackage[T1]{fontenc} 
\usepackage{amssymb}
\usepackage{amsmath}
\usepackage{graphicx}
\title{\boldmath Tachyon inflation in the holographic braneworld}

\author[a,1]{Neven Bili\'c\note{Corresponding author.}}
\author[b]{Dragoljub D.\ Dimitrijevi\'c}
\author[b,1]{Goran S.\ Djordjevic}
\author[b]{Milan Milo\v sevi\'c}
\author[c]{Marko Stojanovi\'c}

\affiliation[a]{Division of Theoretical Physics, Rudjer Bo\v{s}kovi\'{c} Institute,\\Bijeni\v cka cesta 54, 10000 Zagreb, Croatia}
\affiliation[b]{Department of Physics, Faculty of Sciences and Mathematics, University of Ni\v s,\\Vi\v segradska 33, 18000 Ni\v s, Serbia}
\affiliation[c]{Faculty of Medicine, University of Ni\v s,\\Dr Zoran Djindji\' c Boulevard 81, 18000 Ni\v s, Serbia}

\emailAdd{Nevenko.Bilic@irb.hr}
\emailAdd{ddrag@pmf.ni.ac.rs}
\emailAdd{gorandj@junis.ni.ac.rs}
\emailAdd{mmilan@seenet-mtp.info}
\emailAdd{marko.stojanovic@pmf.edu.rs}

\abstract{A model of tachyon  inflation  is proposed in the framework of holographic cosmology. 
The model is based on a holographic braneworld scenario with a D3-brane 
located at the holographic boundary of an asymptotic ADS$_5$ bulk.
The tachyon field that drives inflation is represented by a DBI action on the brane.
We solve the evolution equations analytically  in the slow-roll regime and solve the exact equations numerically.
We calculate the  inflation  parameters and compare the results with Planck 2018  data. }

\arxivnumber{1809.07216}

\begin{document}
\maketitle
\flushbottom

\section{Introduction}

The inflationary universe scenario
has been generally accepted as a solution to the horizon problem and some other 
related problems of the standard Big Bang cosmology.
The origin of the field that drives inflation is still unknown and is subject to speculations.
Among many models of inflation a popular class comprise tachyon inflation models 
\cite{fairbairn,frolov,shiu1,sami,shiu2,kofman,cline,salamate2018,barbosa2018,dantas2018,steer}.
Tachyon models are of particular interest as in these models inflation 
is driven by the tachyon field originating in M or string theory.
The existence of tachyons in the perturbative spectrum of string theory,  both open and closed, 
indicates that the perturbative vacuum is unstable and that there exists a true vacuum
towards which a tachyon field $\theta$ tends \cite{gibbons}.
The basics  
of this process are represented by an effective field theory model \cite{sen}
with a Lagrangian of the Dirac-Born-Infeld (DBI) form 
\begin{equation}
{\cal{L}} = -\ell^{-4} V(\theta/\ell)
\sqrt{1-g^{\mu\nu}\theta_{,\mu}\theta_{,\nu}}  ,
 \label{eq000}
\end{equation}
where $\ell$ is an appropriate length scale, $\theta$ is a scalar field of dimension of length,
and
\begin{equation}
X=g^{\mu\nu}\theta_\mu\theta_\nu .
 \label{eq0042}
\end{equation}
The dimensionless potential $V$ is a positive function of $\theta$ with a unique local maximum 
at $\theta=0$ and a global minimum at $|\theta|=\infty$ at which $V$ vanishes.

 We plan to study a braneworld inflation model in the framework of  a holographic cosmology
\cite{apostolopoulos,bilic1,bilic5,nojiri}. 
By {\it holographic cosmology} we mean a cosmology based on the effective four-dimensional Einstein equations
on the holographic boundary in the framework of 
anti de Sitter/conformal field theory (AdS/CFT) correspondence.
A connection between AdS/CFT correspondence and cosmology has been studied in a different approach
based on a holographic renormalization group flows 
in quantum field theory \cite{kiritsis2,binetruy}.

As we will argue in the next section, the holographic cosmology has a property that 
the universe evolution starts from a point at which the energy density and cosmological scale
are both finite 
rather then from the usual Big Bang singularity of the standard cosmology.
Then the inflation phase proceeds naturally immediately after $t=0$.
Our model is based on a holographic braneworld scenario
with an effective tachyon field on the brane. 
This paper is a sequel to previous  works
\cite{bilic2,bilic3,bilic4,dimitrijevic} in which we have studied tachyon inflation on a Randall-Sundrum 
type of braneworld.
In the present approach a D3-brane is located at the holographic boundary of an asymptotic ADS$_5$ 
bulk. 
We have improved the analytical calculations of 
\cite{bilic5} in the slow role regime up to the second order in the slow role parameters.
We  solve the evolution equations numerically and  
confront our result with the Planck  data.

The remainder of the paper is organized in  four sections and two appendices.
In the next section, Sec. \ref{holographic},  we describe the tachyon cosmology in 
the framework of a holographic braneworld scenario.
The following section, Sec.\ \ref{inflation}, is devoted to a detailed description of
inflation based on the holographic braneworld scenario with tachyon field 
playing the role of the inflaton. 
Our numerical results and comparison with observations are presented in Sec.\ \ref{numerical}.
In Sec.\ \ref{conclusions} we summarize our results 
and give an outlook for
future research.

\section{Holographic tachyon cosmology}
\label{holographic}
Our aim is to study tachyon inflation in the framework of holographic
cosmology. 
We assume that the holographic braneworld is a spatially flat
FRW universe with line element
\begin{equation}
 ds^2=g_{\mu\nu}dx^\mu dx^\nu=dt^2-a^2(t)(dr^2+r^2 d\Omega^2), 
 \label{eq0012}
\end{equation}
and we employ the holographic Friedmann equations (\ref{eq3105}) and (\ref{eq3106})
derived in appendix \ref{appendix}.
If we set $k=0$ and $\mu=0$, these equations take the form
\begin{equation}
h^2-\frac{1}{4}h^4=\frac{\kappa^2}{3}\ell^4\rho ,
 \label{eq115}
\end{equation}
\begin{equation}
\dot{h}\left(1-\frac{1}{2}h^2\right)=-\frac{\kappa^2}{2} \ell^3(p+\rho),
 \label{eq119}
\end{equation}
where, following Ref.\ \cite{bilic3}, we have introduced a dimensionless expansion rate $h\equiv \ell H$
and the fundamental dimensionless coupling
\begin{equation}
\kappa^2=\frac{8\pi G_{\rm N}}{\ell^2}.
\label{eq102}
\end{equation}

The holographic cosmology has interesting properties. Solving the first
Friedmann equation (\ref{eq115}) as a quadratic equation for $h^2$ we find
\begin{equation}
h^2=2\left(1\pm\sqrt{1-\frac{\kappa^2}{3}\ell^4\rho}\right).
 \label{eq125}
\end{equation}
Now, because we do not want our modified cosmology 
to depart too much from the standard cosmology after the inflation era, 
 we demand that Eq.\ (\ref{eq125})
reduces to the standard Friedmann
equation in the low density limit, i.e., in the limit when
$
\kappa^2\ell^4\rho \ll 1$.  
Clearly, this demand will be met only by the ($-$) sign solution in (\ref{eq125}).
Therefore we stick to the ($-$) sign  
and discard the ($+$) sign solution as unphysical. 
Then, it follows
that the physical range of the Hubble expansion rate is between zero and the 
maximal value $h_{\rm max}= \sqrt2$ corresponding to the maximal energy density  
$\rho_{\rm max}= 3/(\kappa^2\ell^4)$ \cite{bilic1,delcampo}.
Assuming no violation of the weak energy condition
$p+\rho\geq 0$, 
 the expansion rate will, according to (\ref{eq119}), be a monotonously decreasing function of time.
The universe evolution starts from $t=0$ with an initial $h_{\rm i} \leq h_{\rm max}$ 
 with energy density and cosmological scale both finite.
Hence, as already noted by C.~Gao \cite{gao}, in the  modified cosmology described
by the Friedmann equations (\ref{eq115}) and (\ref{eq119})
the Big Bang singularity is avoided!

\subsection{Equations of motion}
Tachyon matter in the holographic braneworld is described by
the Lagrangian (\ref{eq000})
in which the scale $\ell$ can be identified with the AdS curvature radius.
The covariant Hamiltonian associated with (\ref{eq000}) is given by \cite{bilic3}
\begin{equation}
{\cal{H}} =\ell^{-4} V\sqrt{1+\eta^2},
 \label{eq0001}
\end{equation}
where
\begin{equation}
\eta=\ell^{4} V^{-1}\sqrt{g_{\mu\nu}\pi^\mu\pi^\nu}
\label{eq2118}
\end{equation}
and the conjugate momentum $\pi^\mu$ is, as usual, related to $\theta_{,\mu}$ via
\begin{equation}
\pi^\mu=
\frac{\partial{\cal{L}}}{\partial\theta_{,\mu}}  .
\label{eq2119}
\end{equation}
From the covariant Hamilton equations
\begin{equation}
\theta_{,\mu}=\frac{\partial{\cal{H}}}{\partial\pi^\mu},
\quad\quad
{\pi^\mu}_{;\mu}=-\frac{\partial{\cal{H}}}{\partial\theta} ,
 \label{eq0011}
\end{equation}
we obtain two first order differential equations in comoving frame 
\begin{equation}
\dot \theta=\frac{\eta}
{\sqrt{1+\eta^2}},
\label{eq003}
\end{equation}
\begin{equation}
\dot \eta=-\frac{3h\eta}{\ell}
-\frac{V_{,\theta}}{V}
\sqrt{1+\eta^2},
\label{eq004}
\end{equation}
where the subscript $,\theta$ denotes a derivative with respect to $\theta$. 
As usual, the Lagrangian and Hamiltonian are identified  with the pressure and energy density, respectively 
i.e., 
\begin{equation}
 p\equiv \mathcal{L}=-\ell^{-4}V\sqrt{1-X} =-\frac{\ell^{-4}V}{\sqrt{1+\eta^2}},
 \label{eq0081}
\end{equation}
\begin{equation}
 \rho\equiv \mathcal{H}=\frac{\ell^{-4}V}{\sqrt{1-X}}=\ell^{-4}V\sqrt{1+\eta^2},
\label{eq008}
\end{equation}
and $X=\dot{\theta}^2$ in comoving frame.
\subsection{Exponential potential}
\label{exponential}
Consider a potential of the form
\begin{equation}
V=V_0 e^{-\omega |\theta|/\ell} ,
 \label{eq305}
\end{equation}
which has been studied extensively in the literature related to string theory and tachyons
\cite{sami,cline,steer,sen2,rezazadeh,nautiyal}.
The dimensionless parameters $\omega$ and $V_0$ are positive and basically free.
It proves advantageous to redefine the field $\theta \rightarrow \theta-\theta_{\rm i}$
and integrate equations of motion from the initial $\theta_{\rm i}$ 
defined as $\theta_{\rm i}=-\ell\omega^{-1}\ln V_0$ instead of integrating from $\theta=0$.
Then, in the physically relevant domain  $\theta_{\rm i}\leq \theta <\infty$, the potential takes the form 
\begin{equation}
 V=e^{-\omega(\theta_{\rm i}+ |\theta-\theta_{\rm i}|)/\ell}=e^{-\omega \theta/\ell}.
 \label{eq135}
\end{equation}
In this way we have traded an arbitrary maximal value $V_0>0$ of the potential at the origin  for
an arbitrary initial value $-\infty <\theta_{\rm i}< \infty$ of the field.
However, as we will discuss next, the initial value $\theta_{\rm i}$ although arbitrary, will be fixed by choosing
initial value for $h$.

The potential (\ref{eq135}) has a convenient property that
one can eliminate dependence on the fundamental parameter $\kappa$ from the equations of motion.
To demonstrate this
we introduce a dimensionless time variable $\tilde{t}=t/\ell$ and replace 
the function $\theta$ by a dimensionless function $y$
defined as
\begin{equation}
y=\frac{\kappa^2}{3}e^{-\omega\theta/\ell}
 \label{eq401}
\end{equation}
Then, from  (\ref{eq003}) and (\ref{eq004}) with (\ref{eq125}) and (\ref{eq008}), we obtain
the following equations of motion
\begin{equation}
\frac{dy}{d\tilde{t}}=- \frac{\omega y\eta}{\sqrt{1+\eta^2}},
 \label{eq402}
\end{equation}
\begin{equation}
\frac{d\eta}{d\tilde{t}}=-3 \eta\left(2-2\sqrt{1-y\sqrt{1+\eta^2}}\right)^{1/2}+\omega\sqrt{1+\eta^2},
 \label{eq403}
\end{equation}
with no $\kappa$-dependence.

\subsection{Remarks on initial conditions}
\label{initial}
To solve equations (\ref{eq003}) and  (\ref{eq004}), or  equations (\ref{eq402}) and (\ref{eq403}), numerically one has to
fix initial values of the functions $\theta$ (or $y$) and $\eta$ at an initial time.
We will assume that the evolution starts at $t=0$ with a given initial expansion rate $h_{\rm i}\leq\sqrt{2}$.
Then, for a chosen initial $\eta_{\rm i}$, the initial $\theta_{\rm i}$ (or $y_{\rm i}$) will be fixed by
the first Friedmann  equation. 
We will  seek solutions imposing either of the two natural initial conditions:  a) $\eta_{\rm i}=0$ or 
b) $\dot{\eta}_{\rm i}=0$. As we shall shortly see, the condition a) assures 
a finite initial $\dot{h}$  whereas  b) 
yields solutions consistent with the slow-roll regime which will be discussed in the next section.
\subsubsection*{a) $\eta_{\rm i}=0$}
In this case  from (\ref{eq0081}) and (\ref{eq008}) it follows 
\begin{equation}
p_{\rm i}=-\rho_{\rm i}
\label{eq131}
\end{equation}
and, as a consequence of (\ref{eq119}),  $\dot{h}_{\rm i}$ will be finite even for 
$h_{\rm i}=\sqrt2$.
The initial $\theta_{\rm i}$ is fixed from (\ref{eq115}) and (\ref{eq008})
\begin{equation}
V(\theta_{\rm i})=\frac{3}{\kappa^2}\left(h_{\rm i}^2-\frac{h_{\rm i}^4}{4}\right).
\end{equation}
For example, the exponential potential (\ref{eq135})
yields
\begin{equation}
\theta_{\rm i}=-\frac{\ell}{\omega}\ln\left[\frac{3}{\kappa^2}\left(h_{\rm i}^2-\frac{h_{\rm i}^4}{4}\right)\right], 
\label{eq130}
\end{equation}
which corresponds to the initial 
\begin{equation}
y_{\rm i}=h_{\rm i}^2-\frac{h_{\rm i}^4}{4} , 
\label{eq136}
\end{equation}
independent of $\kappa$.

\subsubsection*{b) $\dot{\eta}_{\rm i}=0$}
In this case from (\ref{eq004}) it follows 
\begin{equation}
\eta_{\rm i}=-\frac{(\ell V_{,\theta}/V)_{\rm i}}{\sqrt{9 h_{\rm i}^2 - (\ell V_{,\theta}/V)_{\rm i}^2}}
\label{eq132}
\end{equation}
and from (\ref{eq115}) we obtain 
\begin{equation}
 \left(1-\frac{h_{\rm i}^2}{2}\right)^2=1- \frac{\kappa^2}{3}V(\theta_{\rm i})\sqrt{1+\eta_{\rm i}^2} .
\end{equation}
Given $V(\theta)$ these two equations can, in principle,  be solved for $\eta_{\rm i}$ and $\theta_{\rm i}$.
In particular, for the potential (\ref{eq135}) we find
\begin{equation}
\eta_{\rm i}=\frac{\omega}{\sqrt{9 h_{\rm i}^2 - \omega^2}},
\label{eq103}
\end{equation}
 \begin{equation}
\theta_{\rm i}=-\frac{\ell}{\omega}\ln\left[\frac{3}{\kappa^2}
\left(h_{\rm i}^2-\frac{h_{\rm i}^4}{4}\right)\sqrt{1-\frac{\omega^2}{9h_{\rm i}^2}}\right] .
\label{eq133}
\end{equation}
The corresponding 
\begin{equation}
y_{\rm i}=
\left(h_{\rm i}^2-\frac{h_{\rm i}^4}{4}\right)\sqrt{1-\frac{\omega^2}{9h_{\rm i}^2}} 
\label{eq137}
\end{equation}
is again $\kappa$-independent. Because of (\ref{eq103}) the free parameter $\omega$ is restricted by 
$0<\omega < 3\sqrt2$ and the initial $h_{\rm i}$ by $h_{\rm i}> \omega/3$.
Note that  $\eta_{\rm i}$ is always  positive non-zero and hence,
$p_{\rm i}+\rho_{\rm i}>0$. Then, according to (\ref{eq119}), 
for the maximal $h_{\rm i}=\sqrt2$  
the expansion starts with a negative infinite $\dot{h}_{\rm i}$.

\section{Inflation on the holographic brane}
\label{inflation}

Tachyon inflation is based upon the slow evolution of $\theta$ with the slow-roll
conditions \cite{steer}
\begin{equation}
 \dot{\theta}^2\ll 1, \quad |\ddot{\theta}| \ll 3H \dot{\theta}.
 \label{eq4101}
\end{equation}
In view of (\ref{eq003})  the conditions (\ref{eq4101})
are equivalent to    
\begin{equation}
 \eta \ll 1, \quad |\dot\eta| \ll \frac{3 h}{\ell} \eta ,
 \label{eq1001}
\end{equation}
so that in the slow-roll regime  the factors $(1-\dot{\theta}^2)^{-1/2}=
(1+\eta^2)^{1/2}$ in
(\ref{eq0081}) and (\ref{eq008}) may be omitted.
Then, during inflation we have
\begin{equation}
h^2 \simeq 2(1-\sqrt{1-\kappa^2V/3}).
 \label{eq121}
\end{equation}
Note that the second inequality in (\ref{eq1001}) is consistent with the evolution starting from
the initial $\dot{\eta}_{\rm i}=0$ as discussed in section \ref{initial}.

Combining (\ref{eq003}) and (\ref{eq004}) with (\ref{eq4101}) and (\ref{eq1001}) we find
\begin{equation}
 \dot{\theta}\simeq -\frac{\ell V_{,\theta}}{3hV} , 
\label{eq1008}
\end{equation}
\begin{equation}
 \ddot{\theta} \simeq  \frac{\ell V_{,\theta}\dot{h}}{3V h^2} 
 +\left[\left(\frac{V_{,\theta}}{V}\right)^2-\frac{V_{,\theta\theta}}{V}\right]\frac{\ell\dot{\theta}}{3h}.
\label{eq122}
\end{equation}
As mentioned before, the evolution is constrained by the physical range of the expansion 
rate $0\leq h^2\leq 2$.

The most important quantities that characterize inflation are the slow-roll
inflation parameters $\epsilon_j$ defined recursively \cite{steer,schwarz}
\begin{equation}
 \varepsilon_{j+1}= \frac{\dot{\varepsilon}_j}{H\varepsilon_j} ,
\end{equation}
starting from $\varepsilon_0=H_*/H$ , where $H_*$ is the Hubble rate at some
chosen time. The next two are then given by
\begin{equation}
\varepsilon_1\equiv-\frac{\dot{H}}{H^2}\simeq\frac{4-h^2}{12h^2(2-h^2)} \left(\frac{\ell\, V_{,\theta}}{V}\right)^2,
 \label{eq126}
\end{equation}
\begin{equation}
\varepsilon_2\equiv\frac{\dot{\varepsilon_1}}{H\varepsilon_1}
\simeq 2\varepsilon_1\left(1-\frac{2h^2}{(2-h^2)(4-h^2)}\right)+ 
\frac{2\ell^2}{3h^2}\left[\left(\frac{V_{,\theta}}{V}\right)^2-\frac{V_{,\theta\theta}}{V}\right].
 \label{eq127}
\end{equation}
During inflation $\varepsilon_1 < 1$,  $\varepsilon_2 < 1$ and inflation ends once either of the two
exceeds unity.

In the following we will study the exponential potential
\begin{equation}
 V=e^{-\omega \theta/\ell}, 
 \label{eq301}
\end{equation}
as in (\ref{eq135}). 
As we have shown in section \ref{exponential},
this potential has a remarkable property that the evolution does not depend on the fundamental parameter $\kappa$.
For this potential we have
\begin{equation}
\left(\frac{V_{,\theta}}{V}\right)^2=\frac{V_{,\theta\theta}}{V}=\frac{\omega^2}{\ell^2} ,
 \label{eq116}
\end{equation}
so in this case the last term on the right-hand side of (\ref{eq127}) vanishes.
Note that near and at the end of inflation 
$h^2 \ll 1$
 so that we approximately  have $\varepsilon_2\simeq 2\varepsilon_1$.
Hence, the criteria for the end of inflation will be
$\varepsilon_{2\rm f}=1$. 

For the purpose of calculating the spectral index we will also need
the third  slow roll parameter $\varepsilon_3 $  given by
\begin{equation}
\varepsilon_3\equiv\frac{\dot{\varepsilon_2}}{H\varepsilon_2}
\simeq\varepsilon_2 +\frac{4h^2(8-h^4)}{(2-h^2)(4-h^2)(8-8h^2+h^4)}\varepsilon_1,
\label{eq0128}
\end{equation}
where the second equality holds for the exponential potential in the slow-roll approximation.

Equation (\ref{eq1008}) with (\ref{eq301}) may be easily integrated yielding 
the time as a function of $h$
in the slow-roll regime
\begin{equation}
t=\frac{3\ell}{\omega^2}\left[ 2(h_{\rm i}-h)
+\ln\frac{(2-h_{\rm i})(2+h)}{(2+h_{\rm i})(2-h)}
\right],
 \label{eq302}
\end{equation}
where we have chosen the integration constant  so that $h=h_{\rm i}$  at $t=0$.
In our numerical calculations we treat
the initial value $h_{\rm i}^2$ as a free parameter ranging between 0 and 2.

Another important quantity is the number of e-folds $N$ defined as
\begin{equation}
N\equiv\int_{t_{\rm i}}^{t_{\rm f}} H dt
\simeq -3\int_{\theta_{\rm i}}^{\theta_{\rm f}}\frac{h^2V}{\ell^2 V_{,\theta}} d\theta ,
 \label{eq114}
\end{equation}
where the subscripts i and f denote the beginning and the end of inflation, respectively.
Typically $N\simeq$ 50 - 60
is sufficient to solve the flatness and
horizon problems.
The second equality in (\ref{eq114}) is obtained by making use of  Eq. (\ref{eq1008}).
For the potential (\ref{eq301}) the integral can be easily calculated 
and expressed in terms of elementary functions. With the substitution 
\begin{equation}
x\equiv 1-h^2/2=\sqrt{1-\kappa^2 e^{-\omega \theta/\ell}/3}
 \label{eq123}
\end{equation}
we find 
\begin{equation}
N=\frac{6}{\omega\ell}(\theta_{\rm f}-\theta_{\rm i})-\frac{12}{\omega^2}
\int_{x_{\rm i}}^{x_{\rm f}}\frac{x^2dx}{1-x^2} ,
 \label{eq124}
\end{equation}
where
\begin{equation}
x_{\rm i,f} =1-h_{\rm i,f}^2/2=\sqrt{1-\kappa^2 e^{-\omega \theta_{\rm i,f}/\ell}/3} . 
 \label{113}
\end{equation}
The end-value of $\theta$ is fixed by the condition 
$\varepsilon_{2\rm f}=1$, so that
\begin{equation}
\varepsilon_{2\rm f}\simeq\frac{\omega^2}{\kappa^2V_{\rm f}}
=\frac{\omega^2}{\kappa^2}e^{\omega\theta_{\rm f}/\ell}=1 ,
 \label{eq117}
\end{equation}
yielding
\begin{equation}
 \theta_{\rm f}=\frac{\ell}{\omega} \ln \frac{\kappa^2}{\omega^2}
 \label{eq113}
\end{equation}
and 
\begin{equation}
x_{\rm f}=\sqrt{1-\omega^2/3}.
 \label{eq303}
\end{equation}
Using this in (\ref{eq124}) yields
a functional relationship between the parameter $\omega$, the e-fold number $N$, 
and the initial value  $h_{\rm i}$ 
\begin{equation}
N=\frac{12}{\omega^2}\left[
\sqrt{1-\frac{\omega^2}{3}}
-1+\frac{h_{\rm i}^2}{2}+\ln \left(2-\frac{h_{\rm i}^2}{2}\right)
-\ln\left(1+\sqrt{1-\frac{\omega^2}{3}}   \right)
\right] .
 \label{eq111}
\end{equation}
Expanding the expression in brackets up to the lowest order in $\omega^2$ we find
an approximate expression  
\begin{equation}
N=\frac{12}{\omega^2}\left[\frac{h_{\rm i}^2}{2}
+\ln\left(1-\frac{h_{\rm i}^2}{4}\right)
\right]-1.
 \label{eq112}
\end{equation}
As expected for the potential (\ref{eq301}), neither $N$ nor the slow-roll parameters depend on the parameter $\kappa$.

\section{Numerical calculations}
\label{numerical}

\begin{figure}[t!]
\begin{center}
\includegraphics[width=0.45\textwidth,trim= 0 0cm 0 0cm]{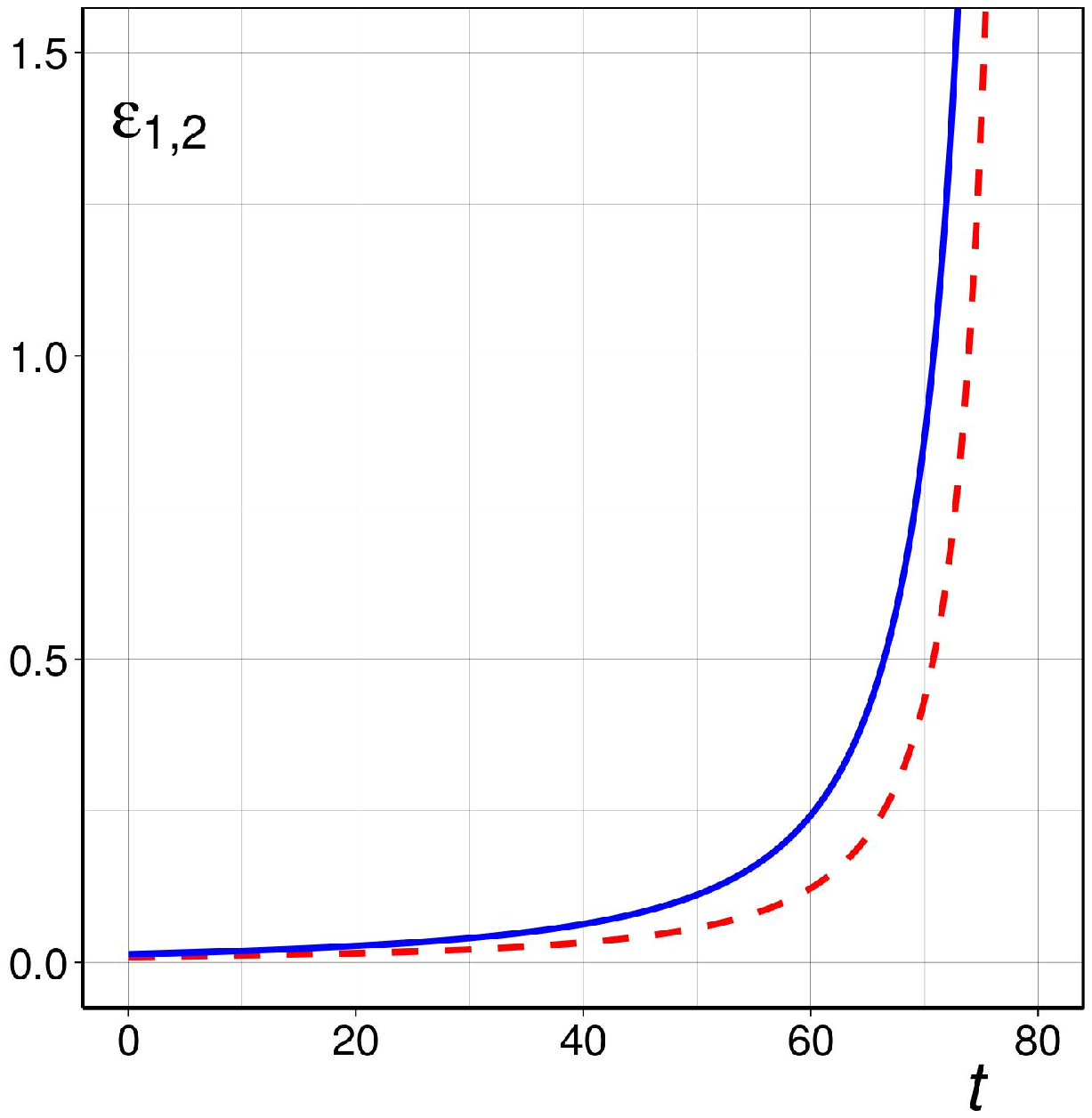}
\hspace{0.02\textwidth}
\includegraphics[width=0.45\textwidth]{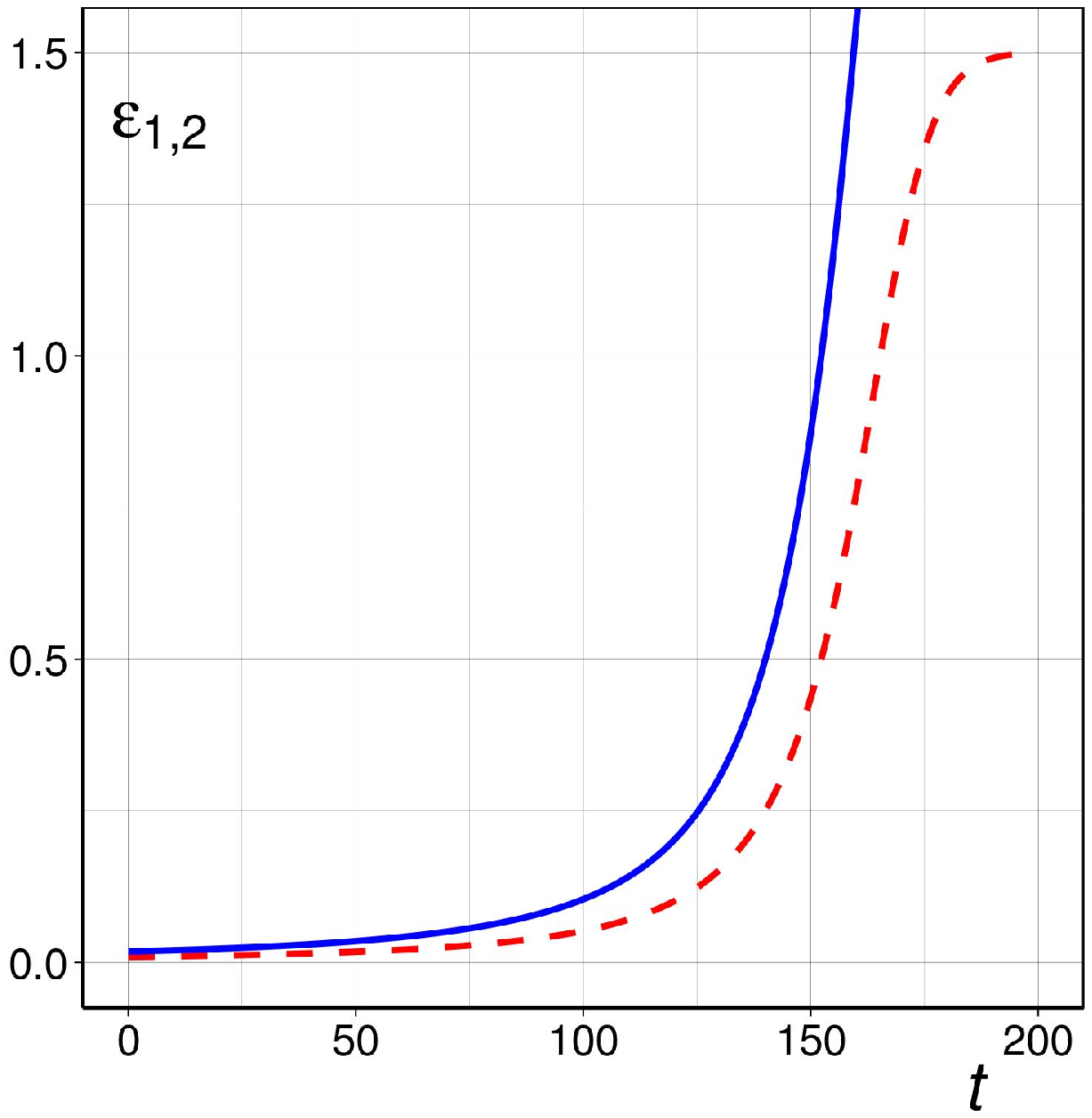}
\caption{Slow-roll parameters $\varepsilon_1$ (dashed red line) and $\varepsilon_2$ (full blue line)
versus time in units of $\ell$ calculated analytically in the slow-roll approximation
(left panel) and numerically (right panel)
for  $\omega^2=0.027$ and  the initial value
$h_{\rm i}^2=0.6$ corresponding to  $N=60$
 according to (\ref{eq111}). 
}
\label{fig1}
\end{center}
\end{figure}

In this section we present the results obtained by numerically solving the exact equations of motion 
(\ref{eq402}) and (\ref{eq403})
given the initial conditions at $t=0$ as  described in 
section \ref{initial}.
The numerical procedure is similar to
that developed in Ref. \cite{bilic2}.
For each pair of randomly chosen $N$ and $h_{\rm i}$ in the intervals  $60\leq N \leq 90$ and
$0< h_{\rm i}^2< 2$, respectively, the parameter $\omega$ is fixed by (\ref{eq112}). 
Then, the set of equations (\ref{eq402}) and  (\ref{eq403}) supplemented by
the equation for $N$
\begin{equation}
 dN=h d\tilde{t}
\end{equation}
is evolved from $t=0$ up to an end time of the order of a few hundreds of $\ell$.
In Fig.\ \ref{fig1} we plot the evolution of the slow roll parameters
for the initial $h_i^2=0.6$ and $\omega^2=0.027$ corresponding to
$N=60$
according to  (\ref{eq111}).
 The inflation actually ends at a time $t_{\rm f}$ 
obtained using  the function $\epsilon_2(t)$  and  demanding $\epsilon_2(t_{\rm f})=1$.
In comparison with the end time obtained analytically in the slow roll approximation
the numerical $t_{\rm f}$ is substantially larger as is evident by comparing
left and right panels of Fig.\ \ref{fig1}.
As a consequence, the numerically calculated $N(t_{\rm f})$ turns out to be larger then the assumed
$N$. Hence, the inflation is assumed to begin at some time $t_{\rm i}>0$, rather than at $t=0$,
such that
\begin{equation}
 N(t_{\rm f})- N(t_{\rm i})=N .
 \end{equation}
The time $t_i$ is then used to find the initial $\epsilon_1(t_i)$ and 
$\epsilon_2(t_i)$ which in turn are  used to calculate the tensor-to-scalar ratio $r$ 
and spectral index $n_{\rm S}$.

\subsection{Spectral index and tensor to scalar ratio}

A proper calculation of the power spectra by perturbing 
the Einstein equations (\ref{eq3002})
would go beyond the scope of the present paper. We propose instead a
simplified scheme described in appendix \ref{perturbations} where we derive approximate expressions
for the scalar and tensor power spectra $\mathcal{P}_{\rm S} $ and $\mathcal{P}_{\rm T}$, respectively.

For scalar perturbations we  calculate the power spectrum 
in the limit when the modes are well outside the acoustic horizon characterized by the comoving wave number
$q=a Hc_{\rm s}^{-1}$.
Here  $c_{\rm s}$ is the adiabatic sound speed
defined by 
\begin{equation}
c_{\rm s}^2\equiv \left.\frac{\partial p}{\partial\rho}\right|_\theta =\frac{p_{,X}}{\rho_{,X}}
=\frac{p+\rho}{2X \rho_{,X}}  ,
\label{eq0018}
\end{equation}
where the subscript $,X$  denotes a derivative with respect to $X$ and
$|_\theta$ means that the derivative is taken keeping $\theta$ fixed, i.e., ignoring the
dependence of $\mathcal{L}$ on $\theta$. This definition coincides with the usual hydrodynamic definition 
of  the sound speed squared as the derivative of pressure with respect to the energy density at fixed
entropy per particle.
For the tachyon fluid described by the Lagrangian (\ref{eq000})
the sound speed squared may be expressed as 
 \begin{equation}
  c_{\rm s}^2=1-X
 =  1-\frac{4(2-h^2)}{3(4-h^2)}\varepsilon_1 ,
  \label{eq3009}
 \end{equation}
where the first equation follows from (\ref{eq0081}) and (\ref{eq008}) and the second equation 
is a consequence of the modified Friedmann equations (\ref{eq115}) and (\ref{eq119}).
This equation shows a substantial deviation from the standard tachyon result \cite{steer}
\begin{equation}
  c_{\rm s}^2|_{\rm st}
 =  1-\frac23\varepsilon_1 .
  \label{eq0067}
 \end{equation}
 The expressions (\ref{eq3009})  and  (\ref{eq0067}) agree
 near the end of inflation, i.e., in the limit $h\rightarrow 0$.
 This is what one would expect since
the holographic cosmology described by equation (\ref{eq125}) reduces to the standard cosmology
in the low density limit.

Next, using the definition (\ref {eq0024}) with (\ref{eq0029}), (\ref{eq0048}), and (\ref{eq0049})
derived in appendix \ref{scalar}, 
  we evaluate the scalar spectral density at the horizon crossing, i.e., for  a wave-number satisfying $q=aH$.
Following Refs.\ \cite{steer,hwang} we make use of the expansion of the Hankel function 
in the limit $c_{\rm s}q\tau\rightarrow 0$  
\begin{equation}
 H_\nu^{(1)}(-c_{\rm s}q\tau)\simeq -\frac{i}{\pi}\Gamma(\nu)\left(\frac{-c_{\rm s}q\tau}{2}\right)^{-\nu},
 \label{eq0053} 
\end{equation}
where $q$ is the comoving wave number and  $\tau$ denotes the conformal time ($\tau<0$).
Using this 
we find  at the lowest order in $\varepsilon_1$ and $\varepsilon_2$
\begin{equation}
 \mathcal{P}_{{\rm S}} \simeq \frac{\kappa^2h^2 }{8\pi^2(1-h^2/2)c_{\rm s}\varepsilon_1 }
 \left[1-2\left(1+C+ \frac{Ch^2}{2-h^2}\right)\varepsilon_1-C \varepsilon_2\right],
\label{eq3007} 
\end{equation}
 where $C=-2+\ln 2 +\gamma\simeq -0.72$ and $\gamma$ is the Euler constant. 
 In comparison with $\mathcal{P}_{{\rm S}}$ obtained in the standard tachyon inflation \cite{steer}, our result is enhanced  
 by a factor $1-h^2/2$ in the denominator on the right side of (\ref{eq3007}) and the linear term
 in $\varepsilon_1$ gets an additional contribution proportional to $h^2/(2-h^2)$.
 
Similarly, from the expression (\ref{eq0065}) with (\ref{eq0069}) for tensor perturbations we obtain
 \begin{equation}
 \mathcal{P}_{\rm T} \simeq \frac{2\kappa^2 h^2}{\pi^2}[1-2(1+C)\varepsilon_1],
 \label{eq3008}
 \end{equation} 
Hence, the tensor perturbation spectrum is given by the usual expression for $\mathcal{P}_{\rm T}$  \cite{steer}.

The scalar spectral index $n_{\rm S}$ and tensor to scalar ratio $r$ are then given by
\begin{equation}
 n_{\rm S}-1= \frac{d\ln \mathcal{P}_{{\rm S}}}{d\ln q},
 \label{eq3006}
 \end{equation}
 \begin{equation}
 r=\frac{\mathcal{P}_{\rm T}}{\mathcal{P}_{{\rm S}}},
 \label{eq3005}
 \end{equation}
 where $\mathcal{P}_{{\rm S}}$ and $\mathcal{P}_{\rm T}$ are 
 evaluated at the horizon crossing.

Using (\ref{eq3007})-(\ref{eq3005}) and
keeping the terms up to the second order in $\varepsilon_i$  we find
\begin{equation}
r=8(2-h^2) \varepsilon_1 \left[1+C\varepsilon_2
+2\left(\frac{Ch^2}{2-h^2}-\frac{2-h^2}{12-3h^2}\right)\varepsilon_1\right] 
 \label{eq128}
\end{equation}
and
\begin{eqnarray}
&& n_{\rm s}=1-\left(2+\frac{2h^2}{2-h^2}\right)\varepsilon_1-\varepsilon_2 
-\left(2+\frac{2h^2}{2-h^2}-\frac{8h^2}{3(4-h^2)^2}-\frac{8C h^2}{(2-h^2)^2}\right)\varepsilon_1^2 
\nonumber\\
&& -\left(\frac83 +\frac{h^2}{3(4-h^2)}+\frac{4C}{2-h^2}\right)\varepsilon_1\varepsilon_2
- C \varepsilon_2\varepsilon_3.
 \label{eq129}
\end{eqnarray}
It is understood that the quantities $h$, $\varepsilon_1$, and $\varepsilon_2$ in these expressions 
are to be taken at the beginning of the slow roll inflation. A comparison of (\ref{eq128}) and (\ref{eq129})  with 
the second  order predictions   
of  the standard tachyon inflation \cite{steer}  
\begin{equation}
r|_{\rm st}=16 \varepsilon_1 (1+C\varepsilon_2-\varepsilon_1/3),
\label{eq138}
\end{equation}
\begin{eqnarray}
 n_{\rm S}|_{\rm st}=1-2\varepsilon_1-\varepsilon_2 
-2\varepsilon_1^2 
-\left(\frac83 +2C\right)\varepsilon_1\varepsilon_2
- C \varepsilon_2\varepsilon_3 
\label{eq139}
\end{eqnarray}
shows a substantial deviation. As expected,
near the end of inflation, i.e., in the limit $h\rightarrow 0$, 
expressions (\ref{eq128}) and (\ref{eq129}) agree with (\ref{eq138}) and (\ref{eq139}), respectively.

\begin{figure}[t!]
\begin{center}
\includegraphics[width=0.7\textwidth,trim= 0 0 0 0]{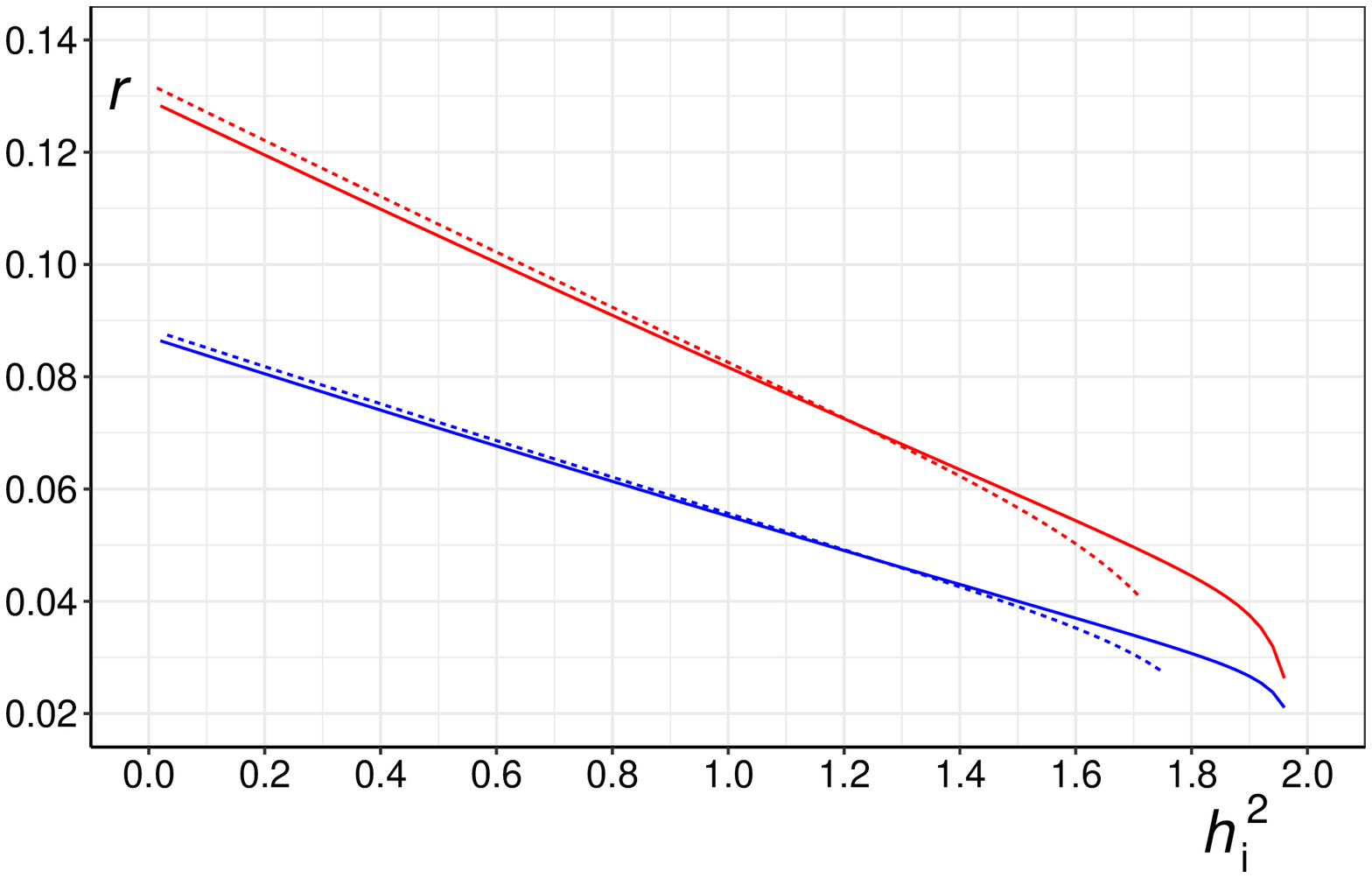}
\caption{$r$ versus initial $h_{\rm i}^2$ calculated analytically (full lines) and numerically
(dashed lines) for fixed $N=60$ (upper red lines) and
$N=90$ (lower blue lines). The parameter $\omega$ is varying along the lines 
in accordance with (\ref{eq112}). 
}
\label{fig2}
\end{center}
\end{figure}
\begin{figure}[t!]
\begin{center}
\includegraphics[width=0.7\textwidth,trim= 0 0 0 0]{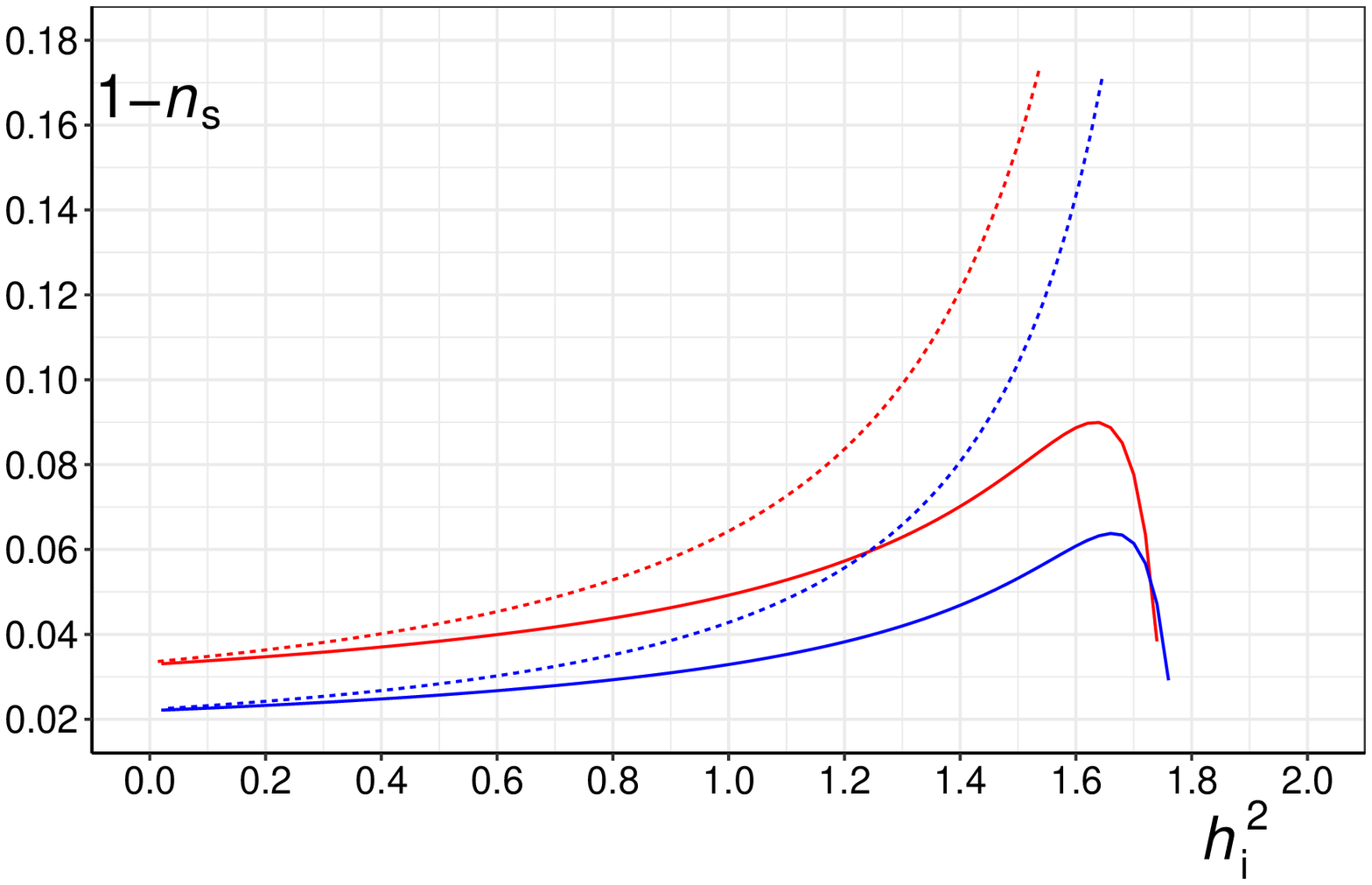}
\caption{$1-n_{\rm S}$ versus initial $h_{\rm i}^2$ calculated analytically (full lines) and numerically
(dashed lines) for fixed $N=60$ (upper red lines) and
$N=90$ (lower blue lines). The parameter $\omega$ is varying along the lines 
in accordance with (\ref{eq112}). 
}
\label{fig3}
\end{center}
\end{figure}

\begin{figure}[t!]
\begin{center}
\includegraphics[width=0.45\textwidth,trim= 0 0cm 0 0cm]{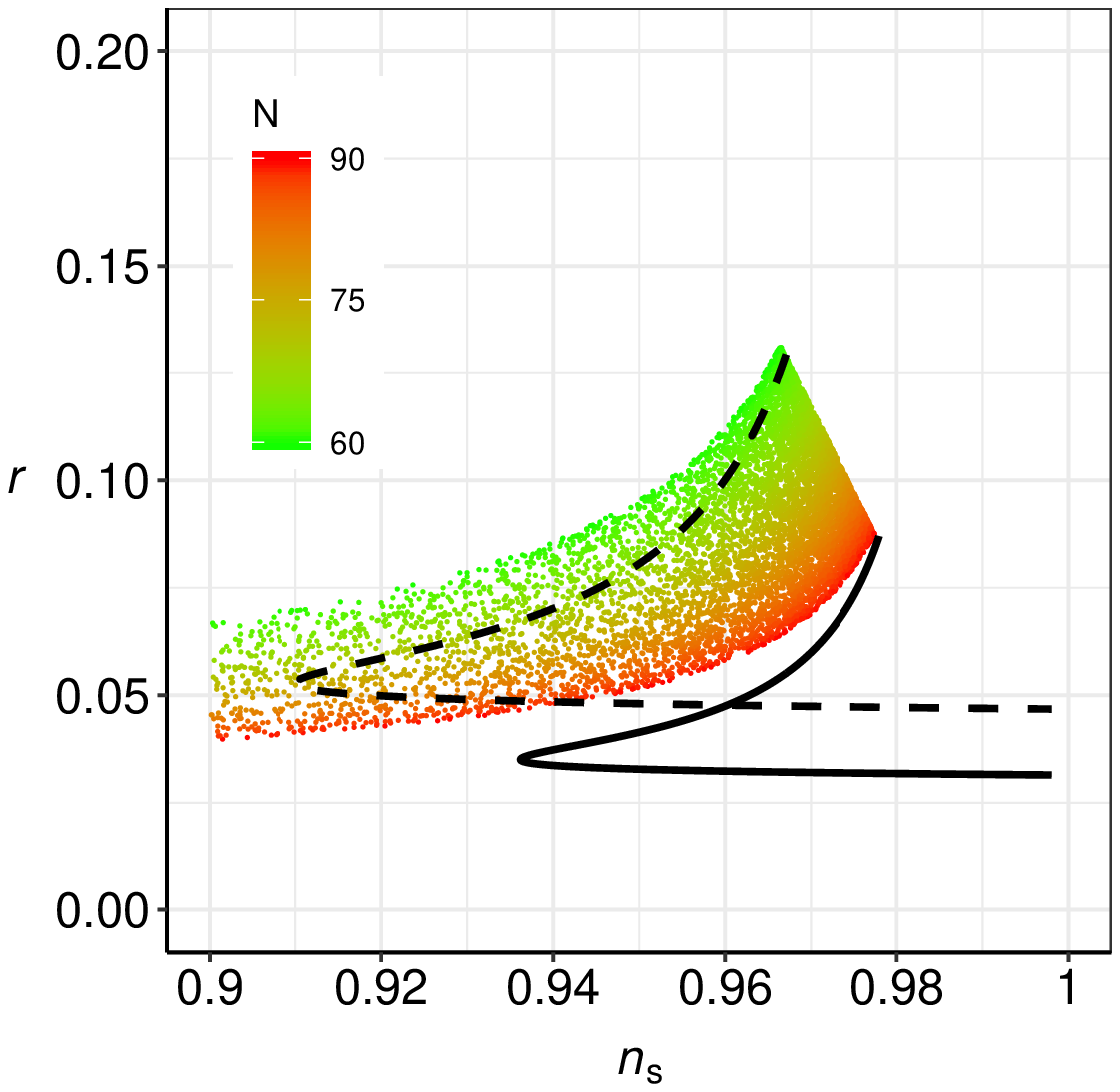}
\hspace{0.02\textwidth}
\includegraphics[width=0.45\textwidth]{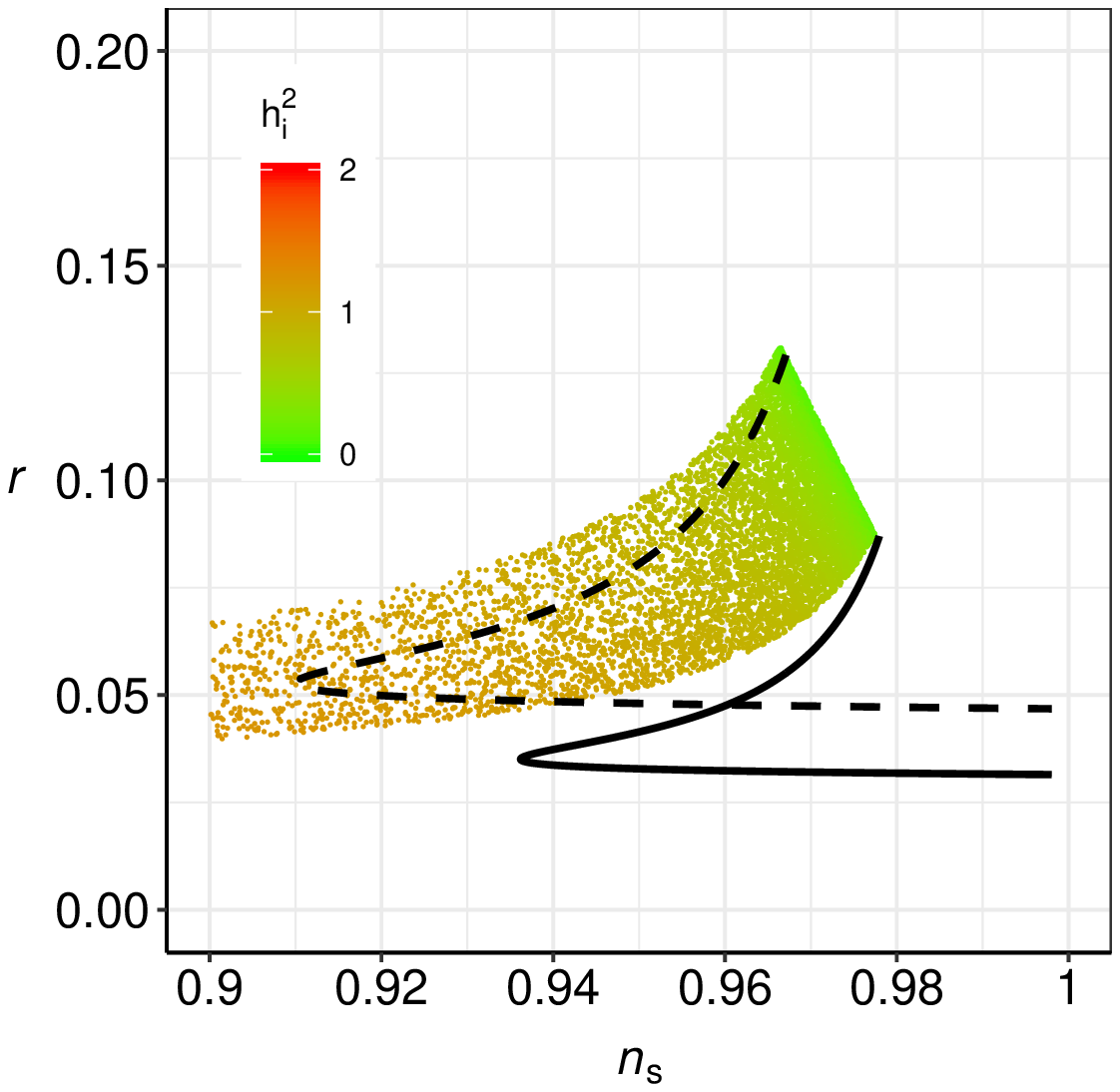}
\caption{$r$ versus $n_{\rm S}$ diagram. 
The dots represent the theoretical predictions 
obtained by solving the equations of motion numerically for randomly chosen 
$N$ ranging between 60 and 90 and  $h_{\rm i}^2$ between 0 and 2.
Variation of color represents  variation of $N$ (left panel)
and variation  of $h_{\rm i}$ (right panel).
The parameter $\omega$  varies in accordance with (\ref{eq112}). 
The analytical results of the slow roll approximation are depicted  by  
the black lines corresponding to  $N=60$ (dashed) and 
  $N=90$ (full).
}
\label{fig4}
\end{center}
\end{figure}

\begin{figure}[t!]
\begin{center}
\includegraphics[width=\textwidth,trim= 0 0 0 0]{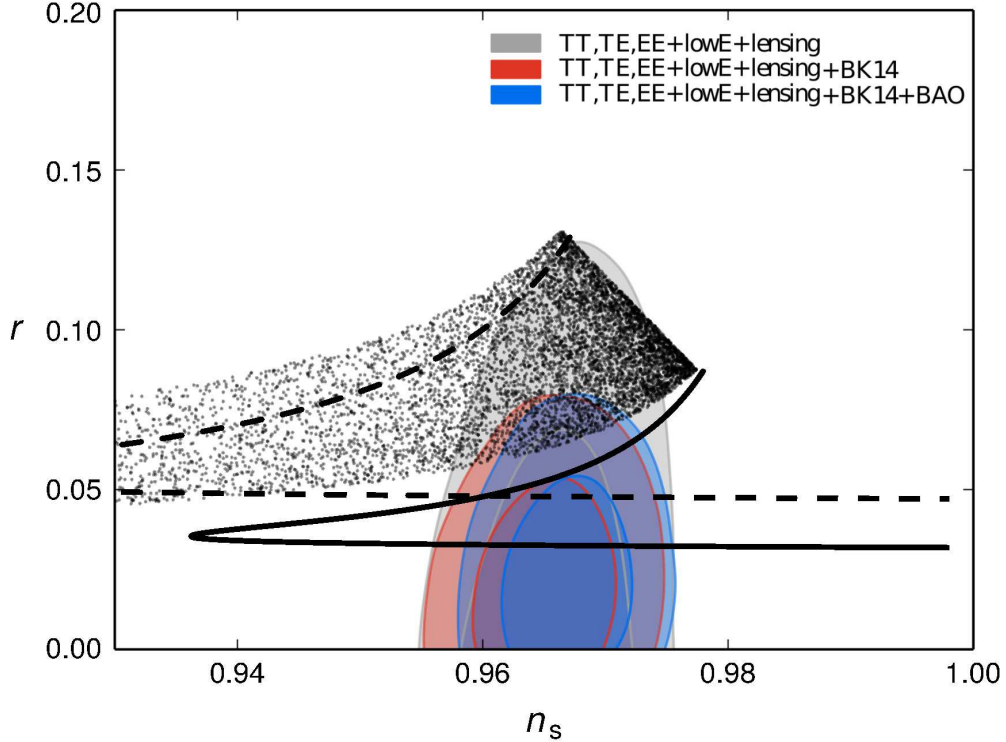}
\caption{$r$ versus $n_{\rm S}$ diagram with observational constraints 
from Ref.\  \cite{planck2018b}. As in Fig.\ \ref{fig4} the dots represent the theoretical predictions 
obtained by solving the equations of motion nu\-me\-ric\-ally for randomly chosen 
$N$  and  $h_{\rm i}^2$ and
the analytical results in the slow roll approximation are depicted  by  
the dashed (N=60) and full (N=90) lines.
}
\label{fig5}
\end{center}
\end{figure}

\begin{figure}[t!]
\begin{center}
\includegraphics[width=\textwidth,trim= 0 0 0 0]{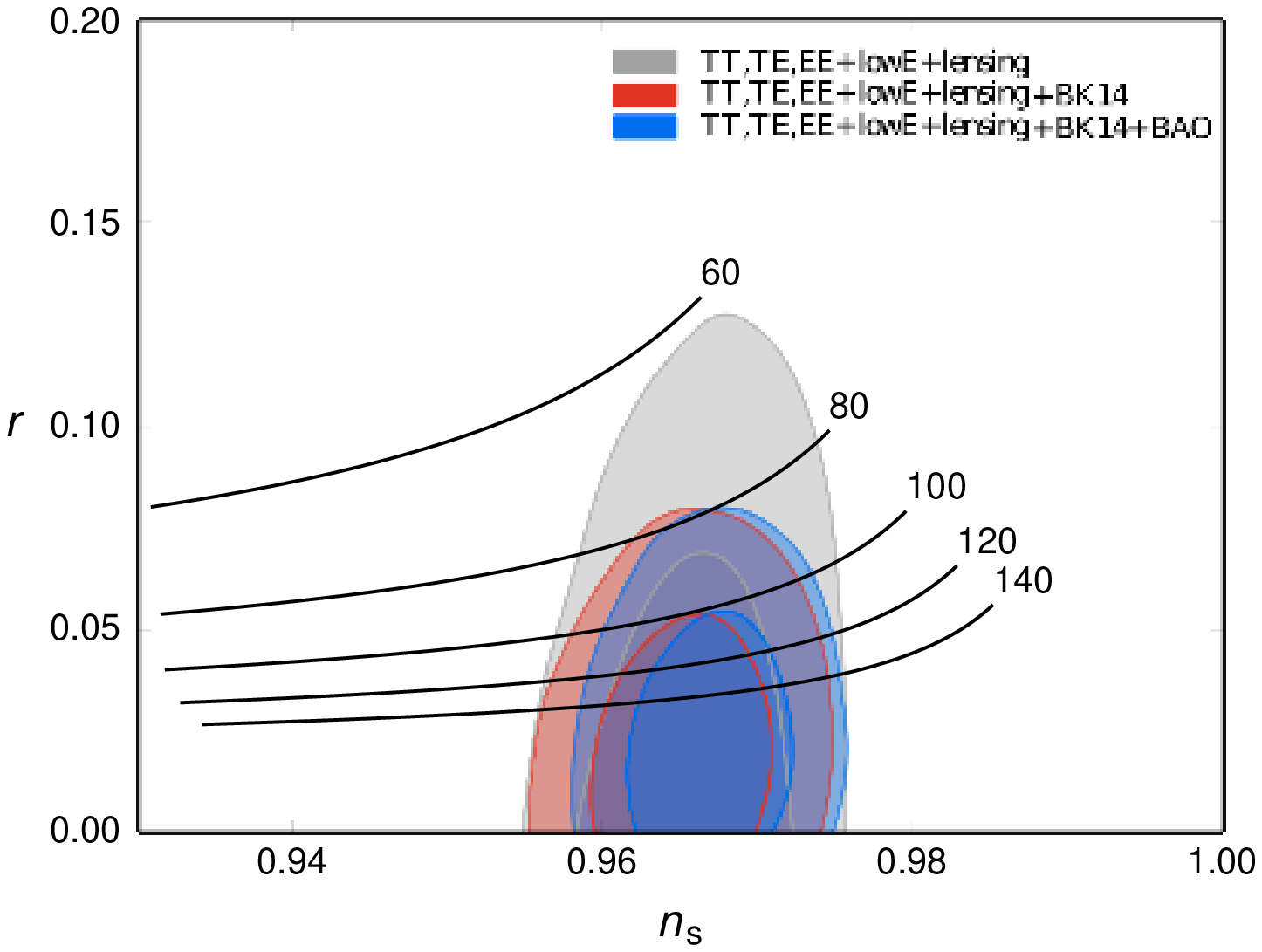}
\caption{$r$ versus $n_{\rm S}$ diagram with observational constraints 
from Ref.\  \cite{planck2018b} for several fixed $N$ (indicated at the end of each line) and varying
$h_{\rm i}^2$  in the interval $0.01-2$.
}
\label{fig6}
\end{center}
\end{figure}

In Figs.\ \ref{fig2} and \ref{fig3} we plot $r$ and $1-n_{\rm S}$ as functions of the initial
Hubble rate squared $h_{\rm i}^2$ for fixed $N$ and varying $\omega$. For each pair $(h_{\rm i}^2,N)$ the parameter 
$\omega$, being in a functional relationship with $N$ and $h_{\rm i}$,
is calculated using the approximate expression
(\ref{eq112}).
 In Fig.~\ref{fig4} we present the $r$ versus  $n_{\rm S}$ diagram.
 The dots represent the numerical data  
for randomly chosen 
$N$ ranging between 60 and 90 and  $h_{\rm i}^2$ between 0 and 2.
Variation of $N$ and $h_{\rm i}^2$ is represented by color.
Each point on the left panel is depicted by a color representing a value of $N$
and similarly on the right panel a value of $h_{\rm i}$.
Clearly, the numerical data set is bounded by the points corresponding to $N=60$ from above and $N=90$ from below.
For each point the parameter $\omega$
is calculated using (\ref{eq112}). 
The results obtained analytically in the slow-roll approximation are depicted by
dashed and full black lines corresponding to  $N=60$ and $N=90$, respectively. 
In Fig.~\ref{fig5} the numerical and analytical data are
superimposed on the observational constraints taken from the Planck Collaboration 2018 \cite{planck2018b}.
 To demonstrate  more explicitly the dependence of  our theoretical predictions on $N$ 
we plot $r$ versus $n_{\rm S}$ in Fig.\ \ref{fig6} for several fixed e-fold numbers $N$
ranging between 60 and 140.
Both numerical and analytical results show that agreement with observations 
is better for larger values of $N$.

 Figs.~\ref{fig4} and \ref{fig5} show that the approximate analytical  results are 
shifted downwards with respect to the numerical data.
This shift reflects the departure of the analytical from the numerical curve in Fig.\ \ref{fig3}
as a consequence of analytical calculations being subject to the slow-roll approximation.
To see this, for definiteness, consider $N=90$. The corresponding numerical results 
are represented in Figs.\ \ref{fig4} and \ref{fig5}
by the points at the lower boundary of the numerical data set and 
the corresponding analytical results 
 by a two-valued function depicted by the full line.
The upper (lower) branch of that curve corresponds to the part of the curve in Fig.\ \ref{fig3} left (right)
from the maximum. The departure of the analytical from the numerical curve in Fig.\ \ref{fig3} 
increases with $h_{\rm i}$ which is consistent with the slow-roll approximation. 
Clearly, the slow-roll approximation breaks down at the maximum of the curve in Fig.\ \ref{fig3} 
so the results represented by the lower branch of the curve in Figs.\ \ref{fig4} and \ref{fig5} 
are not physically relevant.

\subsection{Comment on primordial non-Gaussianity}
\label{nongaussianity}
The prime diagnostic of non-Gaussianity of inflationary fluctuations is described by the three-point
correlation function \cite{baumann}
\begin{equation}
\langle\hat{\zeta}_{q_1}\hat{\zeta}_{q_2}\hat{\zeta}_{q_3}\rangle=
(2\pi)^3 \delta(\mbox{\boldmath $q$}_1+\mbox{\boldmath $q$}_2+\mbox{\boldmath $q$}_3)
f_{\rm NL} F(q_1,q_2,q_3),
 \label{eq0066}
\end{equation}
where $\hat{\zeta}_q$ is the operator associated with the curvature perturbation $\zeta$
introduced in appendix \ref{scalar}.
The quantity $f_{\rm NL}$ is a dimensionless parameter defining the amplitude of non-Gaussianity and
the function $F$ captures the momentum dependence.

Our model belongs to the class of $k$-essence inflation models.
In these models the Lagrangian $\mathcal{L}(\theta,X)$ has a non-canonical dependence on 
the kinetic term $X$ defined in (\ref{eq0042}).
As a consequence, the adiabatic sound speed $c_{\rm s}$ defined by (\ref{eq0018}) in these models
may significantly deviate from 1.
In the $k$-essence models, the largest non-Gaussianity  is peaked at the so called {\em equilateral} configuration
with $q_1\sim q_2\sim q_3$.
The non-Gaussianity amplitude of equilateral triangle $f_{\rm NL}^{\rm equil}$ in a general
$k$-essence is given by \cite{baumann,chen}
\begin{equation}
f_{\rm NL}^{\rm equil} = - \frac{35}{108} \left( \frac{1-c_s^2}{c_s^2} \right) 
+ \frac{20}{81} \Lambda,
 \end{equation}
where
\begin{equation}
 \Lambda \equiv  X^2\frac{ \mathcal{L}_{,XX}^2-
 (1/3)\mathcal{L}_{,X} \mathcal{L}_{,XXX}}{ \mathcal{L}_{,X}^2 + 2 X\mathcal{L}_{,X} \mathcal{L}_{,XX}} .
\end{equation}
Precisely as in the string theory motivated DBI model \cite{silverstein}, 
the quantity $\Lambda$ in the tachyon  model turns out to be identically zero and the amplitude is
directly proportional to $1-c_{\rm s}^2$.
In the tachyon model with standard cosmology one has
\begin{equation}
\left. f_{\rm NL}^{\rm equil}\right|_{\rm st} = - \frac{35}{108} \left( \frac{1-c_s^2}{c_s^2} \right),
\label{eq0068}
 \end{equation}
 where $c_{\rm s}$  is given by (\ref{eq0067}).
 
 Now, we  estimate the non-Gaussianity amplitude in our tachyon model. 
 As the sound speed deviates from unity  most at the end of the slow roll regime,
we estimate the equilateral amplitude at the end of 
inflation neglecting possible post-inflationary effects.
From (\ref{eq0068}) and the calculation of the two point function presented in appendix \ref{scalar},
where the main difference between our model and the standard tachyon inflation is the factor
$1-h^2/2$ which appears  in the denominator of (\ref{eq3007}), we expect  
$f_{\rm NL}^{\rm equil}$ in our model to be of the form (\ref{eq0068}) 
possibly multiplied by a factor $(1-h^2/2)$ raised to some power. 
For the purpose of an estimate this factor can be neglected since $h^2\ll 1$ at the end of the slow roll regime.
An estimate based on (\ref{eq123}), (\ref{eq303}), and $\omega^2=0.027$ yields 
$h^2\lesssim 0.01$ at the end of inflation.
Hence,  we can use the result (\ref{eq0068}) 
with  $c_{\rm s}$  given by (\ref{eq3009}) and $\varepsilon_1=1$ yielding
\begin{equation}
f_{\rm NL}^{\rm equil} \approx  -\frac{70(1-h^2/2)}{108(1+h^2/4)}
\approx \left. f_{\rm NL}^{\rm equil}\right|_{\rm st}= -\frac{70}{108} .
 \label{eq0060}
\end{equation}
 This value is well within the observational constraints provided by the Planck 2015 collaboration \cite{planck2015}:
$ f_{\rm NL}^{\rm equil} =2.6\pm 61.6$ from temperature data and 
$f_{\rm NL}^{\rm equil} =15.6\pm 37.3$ from temperature and polarization data.

In conclusion, the estimated non-Gaussianity in our model at the end of inflationary period 
cannot be distinguished from that in  the standard tachyon inflation.
Possible
post-inflationary persistence of isocurvature perturbations, 
as discussed recently by 
C.\ van de Bruck, T.\ Koivisto, and C.\ Longden \cite{bruck},
may alter this conclusion.  
However, a study of such effects is beyond the scope of the present paper.

 \section{Conclusions}
 \label{conclusions}
We have investigated a model of tachyon inflation based on a holographic braneworld  
scenario with a D3-brane located at the boundary of the ADS$_5$ bulk.
The slow-roll equations in this model turn out to differ substantially  from those
of the standard tachyon inflation with the same potential.
We have studied in particular a simple exponentially attenuating potential.
For a given number of e-folds  our results 
depend only on the initial value of the Hubble rate  and do not depend on the 
fundamental coupling $\kappa$. 
A comparison of our results with  
the most recent observational data \cite{planck2018b,planck2018a} shows reasonable agreement
 as demonstrated in Fig.\ \ref{fig5}.
Apparently, the agreement with observations is better for larger values of 
 the numbers of e-folds $N$. 
 
It would be of considerable interest to 
perform precise calculations for other types of tachyon potentials that are currently 
on the market.
To this end, one would  need to estimate the phenomenologically acceptable range of
the fundamental coupling parameter $\kappa$ and solve the exact
equations numerically for various
potentials. 
Based on our experience from the previous work \cite{bilic2,dimitrijevic} we do not expect a substantial
deviation from the present results.

 \section*{Acknowledgments}

This work has been supported by the ICTP - SEENET-MTP project NT-03 Cosmology - Classical and Quantum Challenges. 
N.~Bili\'c  has been  supported by 
the European Union through the European Regional Development Fund - the Competitiveness and 
 Cohesion Operational Programme (KK.01.1.1.06) and by 
the H2020 CSA Twinning project No.\ 692194, ``RBI-T-WINNING''.
N.~Bili\'c, G.~S.~Djordjevic and M.~Milosevic
are partially supported by the STSM CAN\-TA\-TA-COST grant.
D.~Dimitrijevic, G.~S.~Djordjevic,  M.~Milosevic, and M.~Stojanovic acknowledge
support by the Serbian Ministry of Education, Science and Technological Development 
under the projects No.\ 176021 (G.S.Dj., M.M.), No.\ 174020 (D.D.D.)
and  No.\ 176003 (M.S.).
G.~S.~Djordjevic is grateful to the CERN-TH Department for hospitality and support.

\appendix 

 \section{Cosmology on the holographic brane}
 \label{appendix}
 In this appendix we present a brief review of the  holographic cosmology
 elaborated in \cite{bilic1}.
 The basic idea is to use the action of the second Randall-Sundrum (RSII) model \cite{randall2} 
 as a regulator of the bulk action  and derive the Friedmann  equations
 on the AdS$_5$ boundary using the AdS/CFT prescription and 
holographic renormalization \cite{deharo}.  
 
 A general asymptotically AdS$_5$  metric in Fefferman-Graham coordinates \cite{fefferman}
is of the form
\begin{equation}
ds^2=G_{ab}dx^adx^b =\frac{\ell^2}{z^2}\left( g_{\mu\nu} dx^\mu dx^\nu -dz^2\right),
 \label{eq3001}
\end{equation}
where the length scale $\ell$ is the AdS curvature radius and
we use the Greek alphabet for 3+1 spacetime indices.
Near $z=0$ the metric $g_{\mu\nu}$ can be expanded as
\begin{equation}
  g_{\mu\nu}(z,x)=g^{(0)}_{\mu\nu}(x)+z^2 g^{(2)}_{\mu\nu}(x)+z^4 g^{(4)}_{\mu\nu}(x) +\cdots .
 \end{equation}
 Explicit expressions for $g^{(2n)}_{\mu\nu}$
in terms of arbitrary $g^{(0)}_{\mu\nu}$
can be found in Ref.\ \cite{deharo}.
The pure gravitational on-shell bulk  action is infra-red divergent and can be regularized  
by placing the RSII brane near the
boundary, i.e., at $z=\epsilon \ell$, $\epsilon\ll 1$, so that the induced metric on the brane is
 \begin{equation}
  \gamma_{\mu\nu}=\frac{1}{\epsilon^2}g_{\mu\nu}(\epsilon \ell,x)
=\frac{1}{\epsilon^2}\left(g^{(0)}_{\mu\nu}+\epsilon^2\ell^2 g^{(2)}_{\mu\nu} 
+ \cdots \right) .
 \end{equation}
 The bulk splits in two regions: $0\leq z < \epsilon \ell$ and $\epsilon \ell\leq z < \infty $. We can either
discard the $0\leq z <\epsilon\ell$ region (one-sided regularization) or invoke the Z$_2$
symmetry and identify two regions (two-sided regularization). For
simplicity we shall use the one-sided regularization. The regularized on shell
bulk action is \cite{deharo2}
 \begin{equation} 
S^{\rm reg}[\gamma] =\frac{1}{8\pi G_5} \int\limits_{z\geq \epsilon\ell} d^5x \sqrt{G} 
\left[-\frac{R^{(5)} }{2} -\Lambda _5 \right] 
+S_{\rm GH}[\gamma]+S_{\rm br}[\gamma],
\label{eq001} 
\end{equation}
where $S_{\rm GH}$ is the Gibbons-Hawking boundary  term which is required 
to make a variational procedure well
defined. The brane action is given by
\begin{equation} 
S_{\rm br}[\gamma] =\int d^{4}x\sqrt{-\gamma} (-\sigma + \mathcal{L}[\gamma]),
\label{eq1005}
\end{equation} 
where $\sigma$ is the brane tension and the Lagrangian $\mathcal{L}$ 
describes matter on the brane.
The renormalized  action is obtained by adding
necessary counter-terms and taking the limit $\epsilon\rightarrow 0$
\begin{equation} 
S^{\rm ren}[\gamma]=S^{\rm reg}[\gamma]+S_1[\gamma]+S_2[\gamma]+S_3[\gamma],
\label{eq1007} 
\end{equation}
where the expressions for the counter-terms $S_1$, $S_2$ and $S_3$ may be found in Refs.\ \cite{deharo,hawking}.
Next,  the variation with respect to the induced metric $\gamma_{\mu\nu}$ of
the regularized on shell bulk action (RSII action) should vanish, i.e., we demand
\begin{equation}
 \delta S^{\rm reg}[\gamma]=0 .
\end{equation}
The variation of the action  yields effective four-dimensional Einstein's equations on the
boundary
\begin{equation}
R_{\mu\nu}- \frac12 R g^{(0)}_{\mu\nu}= 8\pi G_{\rm N} 
\left(\langle T^{\rm CFT}_{\mu\nu}\rangle +T_{\mu\nu}
\right),
 \label{eq3002}
\end{equation}
where $R_{\mu\nu}$ is the Ricci tensor associated with the metric $g^{(0)}_{\mu\nu}$
and the energy-momentum tensor 
\begin{equation}
T^{\mu}_{\nu}=\mbox{diag}(\rho, -p,-p,-p) 
\label{eq3010}
\end{equation}
corresponds to the matter Lagrangian $\mathcal{L}$ on the brane.
According to the AdS/CFT prescription, the expectation value of the  energy-momentum
tensor of the dual conformal theory is given by 
\begin{equation}
 \langle T^{\rm CFT}_{\mu\nu}\rangle = \frac{2}{\sqrt{-g^{(0)}}}
 \frac{\partial S^{\rm ren}}{\partial {g^{(0)}}^{\mu\nu} }= 
 \lim_{\epsilon\rightarrow 0} \frac{2}{\sqrt{-g}}
 \frac{\partial S^{\rm ren}}{\partial g^{\mu\nu} }. 
 \label{eq3004}
\end{equation}
This expectation value has been derived explicitly in terms of $g^{(2n)}_{\mu\nu}$, $n=$ 0,1,2, in Ref.\ \cite{deharo}
for an arbitrary metric  $g^{(0)}_{\mu\nu}$ at the $z=0$  boundary.

In the following we will specify the boundary geometry to be of 
a general FRW form 
\begin{equation}
ds_{(0)}^2=g^{(0)}_{\mu\nu}dx^\mu dx^\nu =dt^2 -a^2(t) d\Omega_k^2 ,
 \label{eq3201}
\end{equation}
where 
\begin{equation}
d\Omega^2_k=d\chi^2+\frac{\sin^2(\sqrt{k}\chi)}{k}(d\vartheta^2+\sin^2 \vartheta d\varphi^2)
\label{eq1004}
\end{equation}
is the spatial line element for a 
closed ($k=1$), open hyperbolic ($k=-1$), or open flat ($k=0$) space.
Assuming  an AdS Schwarzschild geometry in the bulk
one obtains \cite{apostolopoulos,bilic1,kiritsis}
\begin{equation}
 \langle T^{\rm CFT}_{\mu\nu}\rangle = t_{\mu\nu}+
\frac14 \langle {T^{\rm CFT}}^\alpha_\alpha\rangle g^{(0)}_{\mu\nu} .
 \label{eq3107}
\end{equation}
The second term on the right-hand side  corresponds to the conformal anomaly 
\begin{equation}
 \langle {T^{\rm CFT}}^\alpha_\alpha\rangle = 
\frac{3\ell^3}{16\pi G_5}\frac{\ddot{a}}{a}\left(H^2+\frac{k}{a^2}\right),
 \label{eq3027}
\end{equation}
where $H=\dot{a}/a$ is the Hubble expansion rate  on  the boundary.
The first term on the right-hand side of (\ref{eq3107}) is
 a traceless tensor, the nonvanishing components of which are
\begin{equation}
 t_{00}=-3 t^i_i =\frac{3\ell^3}{64\pi G_5 }
\left[\left(H^2+\frac{k}{a^2}\right)^2 +\frac{4\mu}{a_0^4} 
-\frac{\ddot{a}_0}{\dot{a}_0}\left(H^2+\frac{k}{a^2}\right)\right],
 \label{eq3108}
\end{equation}
where the dimensionless parameter $\mu$ is related to the black hole mass \cite{myers,witten}.
Then, from (\ref{eq3002}) we obtain the holographic Friedmann equations
\cite{apostolopoulos,bilic1}
\begin{equation}
H^2+\frac{k}{a^2}-\frac{\ell^2}{4}\left(H^2+\frac{k}{a^2}\right)^2=
\frac{8\pi G_{\rm N}}{3}\rho+\frac{\ell^2\mu}{a^4}.
 \label{eq3105}
\end{equation}
From this, by making use of the energy conservation equation
\begin{equation}
\dot{\rho}+3H(p+\rho)=0,
 \label{eq3109}
\end{equation}
we obtain the second Friedmann equation
\begin{equation}
H^2+\frac{k}{a^2}+\frac{\ddot{a}}{a} \left[1-\frac{\ell^2}{2}\left(H^2+\frac{k}{a^2}\right)^2\right]=
\frac{4\pi G_{\rm N}}{3}(\rho-3p),
 \label{eq3106}
\end{equation}
where the pressure $p$ and energy density $\rho$ are the components of  
the energy-momentum tensor as defined in (\ref{eq3010}). 

 \section{Cosmological perturbations}
 \label{perturbations}

  Here we derive the spectra of the cosmological perturbations for the 
 holographic cosmology with tachyon $k$-essence.
  Calculation of the spectra proceeds by identifying the proper canonical field and imposing 
  quantization of the quadratic action for the near free field. 
 The procedure for a general k-inflation is described in  \cite{garriga}
 and applied to the tachyon fluid in Refs.\ \cite{frolov,steer,hwang}.
 
 We shall closely follow J.\ Garriga and V.\ F.\ Mukhanov  \cite{garriga}
 and adjust their formalism to account for the modified Friedmann equations.
In the following we consider a spatially flat background with Friedman equations of the form 
(\ref{eq115}) and (\ref{eq119}) 
in which the pressure and energy  density  $p$ and $\rho$ 
corresponding to the tachyon Lagrangian (\ref{eq000}) are 
defined in (\ref{eq0081}) and (\ref{eq008}).
Equation (\ref{eq115}) can be written in the usual Friedmann form
 \begin{equation}
H^2=\frac{8\pi G_{\rm N}}{3} \tilde{\rho},
\label{eq0009}
\end{equation}
where
 \begin{equation}
\tilde{\rho}=
\frac{6}{\kappa^2\ell^4}\left(1\pm\sqrt{1-\frac{\kappa^2}{3}\ell^4\rho}\right) .
\label{eq0007} 
\end{equation}
Equation (\ref{eq0009}) with (\ref{eq0007}) suggests   
 considering another $k$-essence Lagrangian $\tilde{\mathcal{L}}=\tilde{\mathcal{L}}(X,\theta)$,
 such that the effective energy density $\tilde{\rho}$ defined in  (\ref{eq0007})
 is obtained from $\tilde{\mathcal{L}}$ by the usual prescription
  \begin{equation}
\tilde{\rho}= 2X\tilde{\mathcal{L}}_{,X}-\tilde{\mathcal{L}}.
\label{eq0008}
\end{equation}
Then,
  varying the action
 \begin{equation} 
\tilde{S}=\int d^4x \sqrt{-g} 
\left[-\frac{R}{16\pi G_{\rm N}} +\tilde{\mathcal{L}}(X,\theta) \right] ,
\label{eq0003} 
\end{equation} 
one obtains Einstein's equations
\begin{equation}
R_{\mu\nu}- \frac12 R g_{\mu\nu}= 8\pi G_{\rm N}  
\tilde{T}_{\mu\nu} ,
 \label{eq0004}
\end{equation}
where 
\begin{equation}
\tilde{T}_{\mu\nu}=(\tilde{p}+\tilde{\rho}) u_\mu u_\nu -\tilde{p}g_{\mu\nu}
\label{eq0005}
\end{equation}
and 
\begin{equation}
\tilde{p}\equiv \tilde{\mathcal{L}}, \quad u_\mu =\frac{\theta_{,\mu}}{\sqrt{X}}.
\label{eq0010}
\end{equation}
In principle, the Lagrangian $\tilde{\mathcal{L}}$ can be expressed as an explicit function of
$X$ and $\theta$ by
integrating Eq.\ (\ref{eq0008}) but in the following we will not need
an explicit expression for $\tilde{\mathcal{L}}$.

Assuming isotropy and homogeneity, equations (\ref{eq0004}) yield
the Friedmann equation (\ref{eq0009})
(or equivalently Eq.\ (\ref{eq115})) 
and energy conservation equation 
\begin{equation}
\dot{\tilde{\rho}}+3H(\tilde{p}+\tilde{\rho})=0.
 \label{eq1000}
\end{equation}
In this way  we have obtained the holographic Friedman equation  (\ref{eq3105})
(with $k=0$ and $\mu=0$)
from  a standard $k$-essence action (\ref{eq0003}).
Now, we can apply the procedure of Ref.~\cite{garriga} directly to the modified $k$-essence 
described by (\ref{eq0003}) keeping in mind that the background evolution is governed by our original
equations (\ref{eq003}) and (\ref{eq004}) with (\ref{eq115}) and (\ref{eq119}).


\subsection{Scalar perturbations}
\label{scalar}
Assuming a spatially flat background with line element (\ref{eq0012}), we introduce 
the perturbed line element in the longitudinal gauge
\begin{equation}
 ds^2=(1+2\Phi) dt^2-(1-2\Phi)a^2(t)(dr^2+r^2 d\Omega^2) .
 \label{eq0013}
\end{equation}
Next, we apply directly the procedure of Ref.~\cite{garriga}  to our modified $k$-essence.
The relevant Einstein equations at linear order are given by
\begin{equation}
a^{-2} \Delta \Phi -3H\dot{\Phi} +3H^2\Phi=4\pi G_{\rm N} 
\delta \tilde{T}^0_0 ,
 \label{eq0014}
\end{equation}
\begin{equation}
(\dot{\Phi}+H\Phi)_{,i}=4\pi G_{\rm N} 
\delta \tilde{T}^0_i ,
 \label{eq0015}
\end{equation}
where the  perturbations of  the stress tensor components $\delta \tilde{T}^\mu_\nu$ are induced by the perturbations of
the scalar field $\theta(t,x)=\theta(t)+\delta\theta(t,x)$,
Using the energy conservation (\ref{eq3109}) and the definition (\ref{eq0042}) of $X$ 
one finds 
\begin{equation}
\delta \tilde{T}^0_0=\frac{\tilde{p}+\tilde{\rho}}{\tilde{c}_{\rm s}^2}
\left[\left(\frac{\delta\theta}{\dot{\theta}}\right)^.-\Phi\right] 
-3H(\tilde{p}+\tilde{\rho}) \frac{\delta\theta}{\dot{\theta}} ,
 \label{eq0041}
\end{equation}
\begin{equation}
\delta \tilde{T}^0_i=(\tilde{p}+\tilde{\rho})\left(\frac{\delta\theta}{\dot{\theta}}\right)_{,i} ,
 \label{eq0043}
\end{equation}
where the quantity $\tilde{c}_{\rm s}$ is the adiabatic speed of sound defined by (\ref{eq0018}).
Using (\ref{eq0041}) and (\ref{eq0043}) equations (\ref{eq0014}) and (\ref{eq0015}) 
take the form 
\begin{equation}
  \left(\frac{\delta\theta}{\dot{\theta}}\right)^{\mbox{.}} 
=\Phi + \frac{\tilde{c}_{\rm s}^2}{4\pi G_{\rm N}a^2(\tilde{p}+\tilde{\rho})}\Delta \Phi ,
 \label{eq0033}
\end{equation}
\begin{equation}
(a\Phi)^{\mbox{.}}=4\pi G_{\rm N} a(\tilde{p}+\tilde{\rho})\frac{\delta\theta}{\dot{\theta}}.
 \label{eq0034}
\end{equation}

So far we have merely applied  the formalism of \cite{garriga} in which we have only used the energy conservation
with no need to use the  
modified Friedmann  cosmology so
equations (\ref{eq0033}) and (\ref{eq0034})
coincide with those derived in \cite{garriga}.
However, from now on we invoke the modified Friedmann dynamics 
encoded in equations (\ref{eq115}) and (\ref{eq119}).
That means, in particular, that all background  variables, such as $a$,
$H$, and $\dot{H}$ are obtained by solving equations
(\ref{eq003}) and (\ref{eq004}) with (\ref{eq115}), and,
as may be easily shown, $\tilde{c}_{\rm s}=c_{\rm s}$.
As in Ref.~\cite{garriga}, we introduce  new functions
\begin{equation}
\xi=\frac{a\Phi}{4\pi GH}, \quad
 \zeta = \Phi+H\frac{\delta\theta}{\dot{\theta}}.
\label{eq0021}
\end{equation}
 The quantity $\zeta$ is gauge invariant and  measures the spatial curvature of comoving 
(or constant-$\theta$) hyper-surfaces. During slow-roll inflation $\zeta$ is equal to 
the curvature perturbation on uniform-density hyper-surfaces \cite{baumann}.
Substituting the definitions (\ref{eq0021}) into (\ref{eq0033}) and (\ref{eq0034}) and using 
(\ref{eq119})
we find
\begin{equation}
\dot{\xi}=a\frac{p+\rho}{H^2}\zeta -\frac{h \dot{h}}{2}\xi ,
 \label{eq0044}
\end{equation}
\begin{equation}
\dot{\zeta}= \frac{c_{\rm s}^2 H^2}{a^3 (p+\rho)}\Delta \xi +\frac{\dot{h}}{2}
\left(\zeta-\frac{4\pi G_{\rm N}}{a}H\xi\right).
 \label{eq0045}
\end{equation}
where $h=H\ell$.
Compared with the standard 
equations of Garriga and Mukhanov \cite{garriga}, equations (\ref{eq0044}) and (\ref{eq0045}) 
have additional terms on the righthand sides 
proportional to $\dot{h}$  as a consequence of the 
modified Friedmann equations of the holographic cosmology.

In principle, one can find solutions to these equations numerically.
However, for the sake of comparison with previous calculations in other models,
we prefer to look for approximate solutions in the slow roll regime.
We now show that in this regime the additional terms  in (\ref{eq0044}) and (\ref{eq0045}) 
are suppressed with respect to the standard terms
by a factor  $\varepsilon_1 \equiv -\dot{H}/H^2$.
With hindsight,  we approximate  time derivatives  by
$|\dot{\xi}|\approx H|\xi|$ and $|\dot{\zeta}|\approx H|\zeta|$
and check the validity of this approximation {\em a posteriori}.
With this  we immediately see that the magnitude of the last term 
on the right-hand side of (\ref{eq0044}) is smaller then  
the magnitude of the left-hand side by a factor
$\varepsilon_1 h^2/2$. Then, neglecting the last term  we find an approximate relation
 \begin{equation}
H|\xi|\approx \frac{a (p+\rho)}{H^2}|\zeta|. 
 \label{eq0046}
\end{equation} 
Using this we find that the magnitude of the second term 
on the right-hand side of (\ref{eq0045}) is
of the order $\varepsilon_1 h^2H|\zeta|/2$
and is  suppressed 
with respect to the left-hand side the magnitude of which  is of the order $H|\zeta|$.
For a consistency check we can use 
another  relation 
$|\Delta \xi|\approx q^2|\xi|\approx H^2 a^2 c_{\rm s}^{-2} |\xi|$
approximately valid at the sound horizon crossing.
Using this we
find that the magnitude of the first term on the right-hand side is of the order 
$H|\zeta|$ and it  dominates
the second term and is comparable with the left hand side of (\ref{eq0045}).

By  neglecting the sub-dominant terms and keeping the leading order in $\epsilon_1$, equations 
(\ref{eq0044}) and (\ref{eq0045}) 
can be conveniently expressed as
\begin{equation}
\dot{\xi}=z^2c_{\rm s}^2\zeta,
 \label{eq0016}
\end{equation}
\begin{equation}
\dot{\zeta}=z^{-2}\Delta\xi,
 \label{eq0017}
\end{equation}
where
\begin{equation}
z=\frac{a(p+\rho)^{1/2}}{c_{\rm s}H} = 
\frac{a}{c_{\rm s}}\sqrt{\frac{\epsilon_1}{4\pi G_{\rm N}}\left(1-\frac{h^2}{2}\right)} .
 \label{eq0029}
\end{equation}
Hence, in our approximation we basically neglect the contribution of the conformal fluid 
in the perturbations and the modified Friedman dynamics is reflected 
in a modified  definition of the quantity $z$.

By introducing the conformal time $\tau=\int dt/a$ and a new variable $v=z\zeta$, it is straightforward
to show from equations
(\ref{eq0016}) and (\ref{eq0017}) that  $v$ satisfies a second order differential equation
\begin{equation}
v''-c_{\rm s}^2 \Delta v-\frac{z''}{z}v =0.
 \label{eq0022}
\end{equation}
By making use of the Fourier transformation
\begin{equation}
v(\tau,\mbox{\boldmath $x$})= \frac{1}{(2\pi)^3}\int d^3q  e^{i\mbox{\scriptsize\boldmath $qx$}}v_q(\tau)
 \label{eq0026}
\end{equation}
we also obtain the  mode-function equation
\begin{equation}
v_q''+\left(c_{\rm s}^2q^2  -\frac{z''}{z}\right)v_q =0.
 \label{eq0032}
\end{equation}
As we are looking for a solution to this equation in the slow-roll regime, it is useful to express 
the quantity $z''/z$ in terms of the slow-roll parameters $\varepsilon_i$.
In the slow-roll regime one  can use  the relation 
\begin{equation}
\tau=- \frac{1+\varepsilon_1}{aH}+\mathcal{O}(\varepsilon_i) ,
\label{eq0047}
\end{equation}
which follows from the definition of $\varepsilon_1$ (\ref{eq126}) expressed in terms of the conformal time.
Using this and (\ref{eq0029}) we obtain
at linear order in $\varepsilon_i$ 
\begin{equation}
\frac{z''}{z}=\frac{\nu^2-1/4}{\tau^2},
\label{eq0050}
\end{equation}
where
\begin{equation}
\nu^2=\frac94+\frac32\left(2+\frac{h^2}{2-h^2}\right)
\varepsilon_1+\frac32\varepsilon_2  .
\label{eq0048}
\end{equation}
We look for 
a solution to (\ref{eq0032}) which satisfies the positive frequency asymptotic limit
\begin{equation}
\lim_{\tau\rightarrow -\infty}v_q=\frac{e^{-ic_{\rm s}q\tau}}{\sqrt{2c_{\rm s}q}}.
 \label{eq0027}
\end{equation}
Then the solution which up to a phase agrees with (\ref{eq0027})
is
\begin{equation}
v_q=\frac{\sqrt{\pi}}{2}(-\tau)^{1/2} H_\nu^{(1)}(-c_{\rm s}q\tau),
 \label{eq0049}
\end{equation}
where $H_\nu^{(1)}$ is the Hankel function of the first kind of rank $\nu$.

In the limit of the de Sitter background all $\varepsilon_i$ vanish so $\nu=3/2$ 
in which case the solution to (\ref{eq0032}) with (\ref{eq0050}) is given by
\begin{equation}
v_q=\frac{e^{-ic_{\rm s}q\tau}}{\sqrt{2c_{\rm s}q}}\left(1-\frac{i}{c_{\rm s}q\tau}\right) .
 \label{eq0051}
\end{equation}
Now, we can use this as an approximate solution in the slow-roll regime to check the validity of 
our estimate which led to Eqs.\ (\ref{eq0016}) and  (\ref{eq0017}).
We set $\zeta\approx  v_q/z$  at the 
sound horizon crossing, i.e., we take 
$c_{\rm s}q =aH$. Using Eq.\ (\ref{eq0047}) we find
\begin{equation}
\dot{\zeta}\approx \left(\frac{v_q}{z}\right)^.
\approx \left(\alpha H - \frac{\dot{z}}{z}\right)\zeta ,
 \label{eq0052}
\end{equation}
where $\alpha$ is a complex constant with magnitude of order 1. From (\ref{eq0029}) 
it follows $\dot{z}/z=H + \mathcal{O}(\varepsilon_i^2)$
so $|\dot{\zeta}| \approx H|\zeta|$ in accord with our previously assumed relation.
Then, the relation $|\dot{\xi}|\approx H|\xi|$ also follows 
by virtue of Eqs.\ (\ref{eq0046}) and (\ref{eq0016}).

Next, consider the action 
 for a scalar field  $v$
\begin{equation}
S[v]=\frac12 \int d\tau d^3x \left({v'\,}^2-c_{\rm s}^2(\nabla v)^2+\frac{z''}{z}v^2 \right).
 \label{eq0023}
\end{equation}
The variation of this action obviously yields (\ref{eq0022}) 
 as the equation of motion for $v$.
 Applying the standard canonical quantization \cite{mukhanov}  the field $v_q$ is promoted to an operator
 \begin{equation}
\hat{v}_q=v_q \hat{a}_{\mbox{\scriptsize\boldmath $q$}} +v_{-q}^* \hat{a}^\dagger_{-\mbox{\scriptsize\boldmath $q$}} \, ,
 \label{eq0061}
\end{equation}
where the  operators $\hat{a}_{\mbox{\scriptsize\boldmath $q$}}$  
and $\hat{a}^\dagger_{\mbox{\scriptsize\boldmath $q$}}$ satisfy the canonical commutation relation
 \begin{equation}
[\hat{a}_{\mbox{\scriptsize\boldmath $q$}},\hat{a}^\dagger_{\mbox{\scriptsize\boldmath $q$}'} ]=
(2\pi)^3 \delta(\mbox{\boldmath $q$}-\mbox{\boldmath $q$}') .
 \label{eq0062}
\end{equation} 
Then, the power spectrum of the field $\zeta_q=v_q/z$ is obtained  from the two-point correlation function
\begin{equation}
\langle\hat{\zeta}_q\hat{\zeta}_{q'}\rangle=\langle\hat{v}_q \hat{v}_{q'}\rangle/z^2=
(2\pi)^3 \delta(\mbox{\boldmath $q$}+\mbox{\boldmath $q$}')|\zeta_q|^2.
 \label{eq0063}
\end{equation}
The dimensionless spectral density 
\begin{equation}
\mathcal{P}_{\rm S}(q)=\frac{q^3}{2\pi^2}|\zeta_q|^2=\frac{q^3}{2\pi^2z^2}|v_q|^2 ,
 \label{eq0024}
\end{equation}
with $v_q$ given by (\ref{eq0049}),
characterizes the primordial scalar fluctuations.
The difference with respect to the standard expression is basically in a modified definition of $z$
and in a modification of $v_q$ owing to a new expression (\ref{eq0048}) for the rank $\nu$
of the Hankel function.

 \subsection{Tensor perturbations}
\label{tensor}

 The tensor perturbations are related to the production of gravitational waves
 during inflation. 
 The metric perturbation
 are defined as 
 \begin{equation}
 ds^2= dt^2-a^2(t)\left(\delta_{ij}+h_{ij}\right)dx^idx^j,
 \label{eq0054}
\end{equation}
where $h_{ij}$ is traceless and transverse.
  In the absence of anisotropic stress the gravitational waves are decoupled from matter
and  the relevant Einstein equations at linear order
are
 \begin{equation}
 h_{ij}''+2aH h_{ij}'-\Delta h_{ij}=0.
 \label{eq0055}
\end{equation}
To solve this one uses 
the standard Fourier decomposition 
\begin{equation}
h_{ij}(\tau,\mbox{\boldmath $x$})= \frac{1}{(2\pi)^3}\int d^3q  e^{i\mbox{\scriptsize\boldmath $qx$}}
\sum_s h_q^s(\tau)e_{ij}^s(q),
 \label{eq0056}
\end{equation}
where the polarization tensor
$e_{ij}^s(q)$  satisfies
$q^ie_{ij}^s=0$, and $e_{ij}^s e_{ij}^{s'}=2\delta_{ss'}$
with comoving wave number $q$ and two polarizations $s=+,\times$. 
The amplitude $h_q^s(t)$ then satisfies 
\begin{equation}
 h_q'' +2a H h_q'+q^2h_q=0,
 \label{eq0057}
\end{equation}
where we have suppressed the dependence on $s$ for simplicity and  bear in mind 
that we have to sum over two polarizations in the final expression.
As before, we introduce  a new, canonically normalized variable  
\begin{equation}
v_q=\frac{a}{16\pi G_{\rm N}}h_q
 \label{eq0058}
\end{equation}
which  
satisfies the equation
\begin{equation}
{v_q}''+\left(q^2  -\frac{a''}{a}\right)v_q =0.
 \label{eq0059}
\end{equation}
This equation is of the same form as (\ref{eq0032}) with $c_{\rm s}=1$ and $z$ replaced by $a$.
Then, the properly normalized solution $v_q$ is given by 
\begin{equation}
v_q=\frac{\sqrt{\pi}}{2}(-\tau)^{1/2} H_\nu^{(1)}(-q\tau),
 \label{eq0069}
\end{equation}
with $\nu^2=9/4+3 \varepsilon_1$.
The quantization proceeds in a similar way as in the scalar case
and  the power spectrum of the field $h_q=(16\pi G_{\rm N}/a)v_q$ 
is obtained  from the two-point correlation function
\begin{equation}
\langle\hat{h}_q\hat{h}_{q'}\rangle=\langle\hat{v}_q \hat{v}_{q'}\rangle\frac{\left(16\pi G_{\rm N}\right)^2}{a^2}=
(2\pi)^3 \delta(\mbox{\boldmath $q$}+\mbox{\boldmath $q$}')|h_q|^2.
 \label{eq0064}
\end{equation}
The dimensionless spectral density which characterizes the primordial tensor fluctuations
is given by
\begin{equation}
\mathcal{P}_{\rm T}(q)=\frac{q^3}{\pi^2}|h_q|^2=
\frac{q^3}{\pi^2}\left|\frac{16\pi G_{\rm N}}{a}v_q\right|^2 ,
 \label{eq0065}
\end{equation}
with $v_q$ given by (\ref{eq0069}).


\begin{thebibliography}{1}
\bibitem{fairbairn} 
  M.~Fairbairn and M.~H.~G.~Tytgat,
  {\em Inflation from a tachyon fluid?},
  Phys.\ Lett.\ B {\bf 546}, 1 (2002)
  [hep-th/0204070];
  A.~Feinstein,
  {\em Power law inflation from the rolling tachyon},
  Phys.\ Rev.\ D {\bf 66}, 063511 (2002)
  [hep-th/0204140].
\bibitem{frolov} 
  A.~V.~Frolov, L.~Kofman and A.~A.~Starobinsky,
  {\em Prospects and problems of tachyon matter cosmology},
  Phys.\ Lett.\ B {\bf 545}, 8 (2002)
  [hep-th/0204187];  
\bibitem{shiu1} 
  G.~Shiu and I.~Wasserman,
  {\em Cosmological constraints on tachyon matter},
  Phys.\ Lett.\ B {\bf 541}, 6 (2002)
  [hep-th/0205003]; 
\bibitem{sami} 
  M.~Sami, P.~Chingangbam and T.~Qureshi,
  {\em Aspects of tachyonic inflation with exponential potential},
  Phys.\ Rev.\ D {\bf 66}, 043530 (2002)
  [hep-th/0205179];
\bibitem{shiu2} 
  G.~Shiu, S.~H.~H.~Tye and I.~Wasserman,
  {\em Rolling tachyon in brane world cosmology from superstring field theory},
  Phys.\ Rev.\ D {\bf 67}, 083517 (2003)
  [hep-th/0207119];
  P.~Chingangbam, S.~Panda and A.~Deshamukhya,
  {\em Non-minimally coupled tachyonic inflation in warped string background},
  JHEP {\bf 0502}, 052 (2005)
  [hep-th/0411210];
  S.~del Campo, R.~Herrera and A.~Toloza,
  {\em Tachyon Field in Intermediate Inflation},
  Phys.\ Rev.\ D {\bf 79}, 083507 (2009)
  [arXiv:0904.1032];
  S.~Li and A.~R.~Liddle,
  {\em Observational constraints on tachyon and DBI inflation},
  JCAP {\bf 1403}, 044 (2014)
  [arXiv:1311.4664]. 
%
  \bibitem{kofman} 
  L.~Kofman and A.~D.~Linde,
  {\em Problems with tachyon inflation},
  JHEP {\bf 0207}, 004 (2002)
  [hep-th/0205121].
  \bibitem{cline} 
  J.~M.~Cline, H.~Firouzjahi and P.~Martineau,
  {\em Reheating from tachyon condensation},
  JHEP {\bf 0211}, 041 (2002)
  [hep-th/0207156].
 %
  \bibitem{salamate2018}
  F.~Salamate, I.~Khay, A.~Safsafi, H.~Chakir and M.~Bennai,
  {\em Observational Constraints on the Chaplygin Gas with Inverse Power Law Potential in Braneworld Inflation}, 
  Mosc. Univ. Phys. Bull. {\bf 73} 405 (2018).
%
  \bibitem{barbosa2018}
  N.~Barbosa-Cendejas, R.~Cartas-Fuentevilla, A.~Herrera-Aguilar, R.~R.~Mora-Luna and R.~da Rocha,
  {\em A de Sitter tachyonic braneworld revisited},
  JCAP. {\bf 2018} 005 (2018)
  [hep-th/1709.09016].
%
  \bibitem{dantas2018}
  D.~M.~Dantas, R.~da Rocha and C.~A.~S.~Almeida,
  {\em Monopoles on string-like models and the Coulomb's law},
  Phys. Lett. B. {\bf 782} 149 (2018)
  [hep-th/1802.05638].
%
   \bibitem{steer}
  D.~A.~Steer and F.~Vernizzi,
  {\em Tachyon inflation: Tests and comparison with single scalar field inflation},
  Phys.\ Rev.\ D {\bf 70}, 043527 (2004)
  [hep-th/0310139].
%
\bibitem{gibbons} 
  G.~W.~Gibbons,
  {\em Thoughts on tachyon cosmology},
  Class.\ Quant.\ Grav.\  {\bf 20}, S321 (2003)
  [hep-th/0301117].
%
%
   \bibitem{sen} 
  A.~Sen,
  {\em Supersymmetric world volume action for nonBPS D-branes},
  JHEP {\bf 9910}, 008 (1999)
  [hep-th/9909062].
  
  \bibitem{apostolopoulos} 
  P.~S.~Apostolopoulos, G.~Siopsis and N.~Tetradis,
  {\em Cosmology from an AdS Schwarzschild black hole via holography},
  Phys.\ Rev.\ Lett.\  {\bf 102}, 151301 (2009)
  [arXiv:0809.3505 [hep-th]].
 %
\bibitem{bilic1}
 N.~Bili\'c,
  {\em Randall-Sundrum versus holographic cosmology},
  Phys.\ Rev.\ D {\bf 93}, no. 6, 066010 (2016)
  [arXiv:1511.07323 [gr-qc]].
%
\bibitem{bilic5} 
  N.~Bilic,
  {\em Holographic cosmology and tachyon inflation},
  arXiv:1808.08146 [gr-qc].  
 %
 \bibitem{nojiri} 
  S.~Nojiri and S.~D.~Odintsov,
  Phys.\ Lett.\ B {\bf 494}, 135 (2000)
  [hep-th/0008160].
 %
  \bibitem{kiritsis2} 
  E.~Kiritsis,
  {\em Asymptotic freedom, asymptotic flatness  and cosmology},
  JCAP {\bf 1311}, 011 (2013)
  [arXiv:1307.5873 [hep-th]].
  %
  \bibitem{binetruy} 
  P.~Binetruy, E.~Kiritsis, J.~Mabillard, M.~Pieroni and C.~Rosset,
  {\em Universality classes for models of inflation},
  JCAP {\bf 1504}, no. 04, 033 (2015)
  [arXiv:1407.0820 [astro-ph.CO]].
  %
  \bibitem{bilic2} 
  N.~Bili\'c, D.~Dimitrijevic, G.~Djordjevic and M.~Milosevic,
  {\em Tachyon inflation in an AdS braneworld with backreaction},
  Int.\ J.\ Mod.\ Phys.\ A {\bf 32}, no. 05, 1750039 (2017)
  [arXiv:1607.04524 [gr-qc]].
  \bibitem{bilic3}
N.~Bili\'c, S.~Domazet and G.~S.~Djordjevic,
  {\em Particle creation and reheating in a braneworld inflationary scenario},
  Phys.\ Rev.\ D {\bf 96}, no. 8, 083518 (2017)
  [arXiv:1707.06023 [hep-th]]
     \bibitem{bilic4} 
  N.~Bili\'c, S.~Domazet and G.~Djordjevic,
  {\em Tachyon with an inverse power-law potential in a braneworld cosmology},
  Class.\ Quant.\ Grav.\  {\bf 34}, no. 16, 165006 (2017)
  [arXiv:1704.01072 [gr-qc]]. 
%
\bibitem{dimitrijevic}
  D.~D.~Dimitrijevi\'c, N.~Bili\'c, G.~.S.~Djordjevic, M.~Milo\v sevi\'c, M.~Stojanovi\'c,
  {\em Tachyon scalar field in a braneworld cosmology},
  Int.\ J.\ Mod.\ Phys.\ A {\bf 33}, no. 34, 1845017 (2018)  

\bibitem{delcampo}
 S.~del Campo,
  {\em Approach to exact inflation in modified Friedmann equation},
  JCAP {\bf 1212}, 005 (2012)
  [arXiv:1212.1315 [astro-ph.CO]]. 
%
  \bibitem{gao} 
  C.~Gao,
  {\em Generalized modified gravity with the second order acceleration equation},
  Phys.\ Rev.\ D {\bf 86}, 103512 (2012)
  [arXiv:1208.2790 [gr-qc]].
%
\bibitem{sen2} 
  A.~Sen,
  {\em Field theory of tachyon matter},
  Mod.\ Phys.\ Lett.\ A {\bf 17}, 1797 (2002)
  [hep-th/0204143].
%
\bibitem{rezazadeh} 
  K.~Rezazadeh, K.~Karami and S.~Hashemi,
  {\em Tachyon inflation with steep potentials},
  Phys.\ Rev.\ D {\bf 95}, no. 10, 103506 (2017)
  [arXiv:1508.04760 [gr-qc]].
 %
 \bibitem{nautiyal} 
  A.~Nautiyal,
  {\em Reheating constraints on Tachyon Inflation},
  Phys.\ Rev.\ D {\bf 98}, no. 10, 103531 (2018).
%
\bibitem{schwarz} 
  D.~J.~Schwarz, C.~A.~Terrero-Escalante and A.~A.~Garcia,
  {\em Higher order corrections to primordial spectra from cosmological inflation},
  Phys.\ Lett.\ B {\bf 517}, 243 (2001)
  [astro-ph/0106020].
\bibitem{hwang} 
  J.~c.~Hwang and H.~Noh,
  {\em Cosmological perturbations in a generalized gravity including tachyonic condensation},
  Phys.\ Rev.\ D {\bf 66}, 084009 (2002)
  [hep-th/0206100]. 
  %
 %
 \bibitem{planck2018b} 
  Y.~Akrami {\it et al.} [Planck Collaboration],
  {\em Planck 2018 results. X. Constraints on inflation},
  arXiv:1807.06211 [astro-ph.CO].    
 %
 \bibitem{baumann} 
  D.~Baumann,
{\em Inflation},
  in Proceedings of TASI09, Physics of the Large and the Small, pp. 523-686 (2011).
  arXiv:0907.5424 [hep-th].
    \bibitem{chen} 
  X.~Chen, M.~x.~Huang, S.~Kachru and G.~Shiu,
{\em Observational signatures and non-Gaussianities of general single field inflation},
  JCAP {\bf 0701}, 002 (2007)
  [hep-th/0605045].
  %
  \bibitem{silverstein} 
  E.~Silverstein and D.~Tong,
{\em Scalar speed limits and cosmology: Acceleration from D-cceleration},
  Phys.\ Rev.\ D {\bf 70}, 103505 (2004)
  [hep-th/0310221].  
%

\bibitem{planck2015} %
  P.~A.~R.~Ade {\it et al.} [Planck Collaboration],
{\em Planck 2015 results. XVII. Constraints on primordial non-Gaussianity},
  Astron.\ Astrophys.\  {\bf 594}, A17 (2016)
  [arXiv:1502.01592 [astro-ph.CO]].
%
\bibitem{bruck} 
  C.~van de Bruck, T.~Koivisto and C.~Longden,
 {\em Non-Gaussianity in multi-sound-speed disformally coupled inflation},
  JCAP {\bf 1702}, no. 02, 029 (2017)
  [arXiv:1608.08801 [astro-ph.CO]].  
  
\bibitem{planck2018a} 
  N.~Aghanim {\it et al.} [Planck Collaboration],
  {\em Planck 2018 results. VI. Cosmological parameters},
  arXiv:1807.06209 [astro-ph.CO]. 


 
%

   
%


  \bibitem{randall2}
L.~Randall and R.~Sundrum, 
{\em An Alternative to Compactification},
Phys.\ Rev.\ Lett. {\bf 83}, 4690 (1999).
%
%
%
%
\bibitem{deharo} 
  S.~de Haro, S.~N.~Solodukhin and K.~Skenderis,
  {\em Holographic reconstruction of space-time and renormalization in the AdS / CFT correspondence},
  Commun.\ Math.\ Phys.\  {\bf 217}, 595 (2001)
  [hep-th/0002230].
 %
\bibitem{fefferman}
C.~Fefferman and C.~R.~Graham,
{\em The ambient metric}
arXiv:0710.0919 [math.DG]
 %
 \bibitem{deharo2} 
  S.~de Haro, K.~Skenderis and S.~N.~Solodukhin,
  {\em Gravity in warped compactifications and the holographic stress tensor},
  Class.\ Quant.\ Grav.\  {\bf 18}, 3171 (2001)
  [hep-th/0011230].
%
 \bibitem{hawking} 
  S.~W.~Hawking, T.~Hertog and H.~S.~Reall,
  {\em Brane new world},
  Phys.\ Rev.\ D {\bf 62}, 043501 (2000)
  [hep-th/0003052].
%
  \bibitem{kiritsis} 
  E.~Kiritsis,
  {\em Holography and brane-bulk energy exchange},
  JCAP {\bf 0510}, 014 (2005)
  [hep-th/0504219].
%
  \bibitem{myers} 
  R.~C.~Myers and M.~J.~Perry,
  {\em Black Holes in Higher Dimensional Space-Times},
  Annals Phys.\  {\bf 172}, 304 (1986).
%
  \bibitem{witten} 
  E.~Witten,
  {\em Anti-de Sitter space, thermal phase transition, and confinement in gauge theories},
  Adv.\ Theor.\ Math.\ Phys.\  {\bf 2}, 505 (1998)
  [hep-th/9803131]. 

\bibitem{garriga} 
  J.~Garriga and V.~F.~Mukhanov,
  {\em Perturbations in k-inflation},
  Phys.\ Lett.\ B {\bf 458}, 219 (1999)
  [hep-th/9904176].

 \bibitem{mukhanov} 
  V.~F.~Mukhanov, H.~A.~Feldman and R.~H.~Brandenberger,
 {\em Theory of cosmological perturbations. Part 1. Classical perturbations. Part 2. 
 Quantum theory of perturbations. Part 3. Extensions},
  Phys.\ Rept.\  {\bf 215}, 203 (1992).



\end{thebibliography}
\end{document}